\documentclass{aa}  

\usepackage{txfonts}
\usepackage[pdfpagelabels=false,colorlinks=true,linkcolor=blue,citecolor=blue,filecolor=blue,urlcolor=blue]{hyperref}
\usepackage{natbib}
\bibpunct{(}{)}{;}{a}{}{,}
\usepackage[utf8]{inputenc}

\usepackage{amsmath}
\usepackage{graphicx}
\usepackage{float}
\usepackage{placeins}

\usepackage{color}
\usepackage{xcolor}
\usepackage{soul}
\newcommand{\lbrac}[1]{\left(#1\right)}
\newcommand{\comm}[1]{{}}
\newcommand{\ftides}{f_{\rm tides}}
\newcommand{\fcomp}{f_{\rm conv}}

\newcommand{\fstreams}{f_{\rm str}}
\newcommand{\fstr}{f_{\rm str}}
\newcommand{\kpc}{{\rm kpc}}
\newcommand{\Msun}{{\rm M_\odot}}
\newcommand{\mean}[1]{\left<#1\right>}
\newcommand{\pc}{{\rm pc}}
\newcommand{\shear}
{\overleftrightarrow{S}}
\newcommand{\curl}[1]{\nabla\times{#1}}
\newcommand{\divr}[1]{\nabla\cdot{#1}}
\newcommand{\partd}[2]{\frac{\partial #1}{\partial #2}}
\defcitealias{Mandelker24}{M25}
\newcommand{\mycitet}[1]{\citeauthor{#1}\ \citeyear{#1}}
\begin{document}

   \title{On the origin of compressive turbulence in protoclumps in high redshift disks}

   \author{Omry Ginzburg
          \inst{1}
          \and
          Avishai Dekel\inst{1,2}
          \and
          Nir Mandelker\inst{1}
          \and
          Dhruba Dutta Chowdhury\inst{1}
          \and
          Frederic Bournaud\inst{3}
          \and
          Daniel Ceverino\inst{4,5}
          \and
          Joel Primack\inst{6}
            }
   \institute{Racah Institute of Physics, The Hebrew University, Jerusalem 91904 Israel
         \and
             SCIPP, University of California, Santa Cruz, CA 95064, USA
         \and 
         AIM, CEA, CNRS, Université Paris-Saclay, Université Paris Diderot, Sorbonne Paris Cité, 91191 Gif-sur-Yvette, France
         \and
         Departamento de Fisica Teorica, Modulo 8, Facultad de Ciencias, Universidad Autonoma de Madrid, 28049 Madrid, Spain
         \and
         CIAFF, Facultad de Ciencias, Universidad Autonoma de Madrid, 28049 Madrid, Spain
         \and
         Department of Physics, University of California, Santa Cruz, CA, 95064, USA
             }
   \date{Received -; accepted -}

  \abstract
   {The giant, star forming clumps in gas-rich, high redshift   disks are commonly  assumed to form due to gravitational instabilities, in which protoclumps have a Toomre-$Q$ parameter less than unity. However, some cosmological simulations have shown that clumps can form in regions where $Q$ is significantly greater than unity. In these simulations, there is an energy excess of compressive modes of turbulence that lead to gravitational collapse of regions that were not supposed to collapse under their own self-gravity, according to linear Toomre instability. In contrast, sites of clump formation in isolated simulations do not show this excess, hinting that the origin of the compressive turbulence may be external.}
   {We explore two external mechanisms that can induce the compressive modes of disk turbulence in protoclumps, namely, compressive tides exerted by the cosmological environment and the direct driving by inflowing streams. }
   {We correlate the local strength of compressive tides and the amount of fresh stream material with protoclump regions in zoom-in cosmological simulations. The local strength of compressive tides is derived from the eigenvalues of the tidal tensor. The local strength of incoming streams is derived from the fractional presence of the stream compared to the average.}
   {We find that the tidal field in protoclumps tends to be over-compressive while random patches in the disk show substantial diverging tides.  In particular, in $25\%$ of the protoclumps, the tidal field is fully compressive, while no random patch resides in regions of fully compressive tides. In addition, the protoclumps tend to reside in regions where the fraction of incoming stream mass is 2-10 times larger than the average at the same galactocentric radius.}
   {Both compressive tides and inflowing streams are correlated with the protoclumps and can thus serve as the drivers of excessive compressive turbulence that can initiate clump formation before self-gravity takes over. This constitutes a new, non-linear mode of violent disk instabilities in high-$z$ galaxies.}

   \keywords{ ISM: Kinematics and dynamics -- ISM: Structure
                 -- Galaxies: formation -- Galaxies: high-redshift
               }

   \maketitle
%

\section{Introduction}
High-redshift galactic disks are observed to be highly perturbed, sustaining supersonic levels of turbulence \citep{Hennebelle12,Kassin12,Rowland24}, and hosting giant, kiloparsec-scale clumps \citep{Genzel2008, G18, Zanella19, Fujimoto24}. Approximately $40-60\%$ of galaxies in the redshift range $1 \leq z \leq 4$ appear to exhibit such clumpy structures \citep{G15,Shibuya2016}. These clumps are prominent across various wavelengths, appearing in rest-frame UV \citep{Wuyts2012, G15, Sattari23}, ${\rm H\alpha}$ emission \citep{Genzel11, Swinbank12, Livermore15}, ${\rm CII}$ \citep{Zanella24}, as well as rest-frame optical and IR \citep{Forster2011,Kalita24}. They contribute roughly $10-20\%$ of the galaxy’s total star formation rate (SFR), while their observed masses are more uncertain due to the limited resolution \citep{Cava18,Meng2020,HuertasCompany20}.

The presence of clumps in galactic disks significantly influences their structural and dynamical evolution. Clumps are believed to play a pivotal role in radial mass transport within the disk (\mycitet{DSC}; \mycitet{DekelBurkert14}; \mycitet{Genzel23}; cf. \mycitet{DuttaChowdhury24}), in contributing to bulge growth \citep{Zolotov15,Lapiner23}; in creating cored dark matter halos through dynamical friction heating \citep{Ogiya22}; and in driving turbulence via clump-clump and clump-disk interactions \citep{DSC,KB10}. These processes are relevant if clumps persist long enough to participate in them. Clump longevity remains under debate, with some studies suggesting a substantial fraction of long-lived clumps \citep{Mandelker17} and others arguing for a prolonged clumpy phase composed of short-lived clumps \citep{Oklopcic2017}. Disk properties, such as gas fraction, contribute to clump longevity \citep{Renaud24}, but stellar and supernova feedback are likely the primary factors. Thus, clumps hold critical insights into feedback processes in galaxies (\mycitet{Mayer2016}; \mycitet{Ceverino23}; \mycitet{DekelMigration}; cf. \mycitet{Fensch2020}), helping to constrain them.

Observationally, clump age gradients support the existence of long-lived clumps and their migration toward the galactic center \citep{Shibuya2016,G18}. In \citet{Ginzburg21}, we applied deep learning techniques to identify clumps in star-forming galaxies from the CANDELS survey, and classify them by longevity (short- or long-lived), based on cosmological simulations of clumpy galaxies \citep{Mandelker17}. Our findings indicated that these galaxies host long-lived clumps that tend to migrate inward. Theoretically, analytical models of clump survival and migration suggest that clumps in gas-rich galaxies with stellar masses around $M_{*}\!\sim\!10^{9.3}\,\Msun$ at $z\!\sim\!2$ are likely to survive and migrate, typically reaching the center within $\sim\!10$ clump free-fall times \citep{DekelMigration}. Cosmological simulations, however, present a more complex picture. While the VELA3 simulations \citep{Ceverino14,Zolotov15,Mandelker17} clearly indicate the presence of long-lived clumps, the VELA6 simulations \citep{Ceverino23}, as well as the FIRE simulations \citep{Oklopcic2017} predominantly show short-lived clumps. The FIRE simulations typically exhibit outflow mass-loading factors $\eta\equiv\dot{M}_{\rm out}/{\rm SFR} \sim 4-10$ and outflow velocities of $400-700\ {\rm km\ s^{-1}}$ \citep{Muratov2015} in the relevant mass and redshift range. In contrast, for massive, long-lived clumps in VELA (as seen in Fig. 13 of \citet{Mandelker17}), the mass-loading factor is around unity or less. This mass-loading factor is consistent with observational estimate for ionized gas outflows driven by stellar and supernova feedback \citep{Forster2019}.

While the giant clumps could have formed ex situ, being the remnants of small galaxies that merged with the disk, it is believed that most clumps are formed in situ, due to local gravitational collapse. The common understanding of in situ clump formation is that they arise from violent disk instability, governed by the Toomre instability \citep{Toomre64}. In this picture, a razor thin disk becomes linearly unstable to axisymmetric perturbations when $Q\equiv \kappa\sigma/\pi G\Sigma<1$. Here, $\kappa$ is the epicyclic frequency, $\sigma$ is the radial velocity dispersion (either thermal or turbulent), and $\Sigma$ is the surface density. While it is convenient and quite common to evaluate the $Q$ parameter using global values of the disk, the $Q$ parameter is derived locally, at each galactocentric radius, by assuming the scales of the perturbations are much smaller than the disk scale radius \citep{BT08}. {When taking the disk's thickness into account, \citet{RomeoFalstad2013}, following \citet{Romeo1994}, derived an order unity correction factor to the two dimensional $Q$ parameter that depends on the anisotropy of the turbulence velocity field. The correction factor increases the $Q$ parameter, which has a stabilizing effect on the disk -- an increased $Q$ parameter is equivalent to a decreased threshold for stability, to a typical value of $\!\sim\!0.68$ for thick galactic disks \citep{GoldreichLyndenBell}.}

Disks are typically assumed to be in a state of “marginal instability” where {$Q\!\sim\!Q_{\rm crit}$, where $0.68\!\lesssim Q_{\rm crit}\!\lesssim2$ (see below) the threshold for instability}. This is based on the following qualitative argument \citep{Noguchi99,DSC}: {if a disk’s $Q$ drops below $Q_{\rm crit}$}, the disk starts to fragment into rings in the linear regime, which later fragment non-linearly into {filaments and feathers \citep{Arora25} which eventually collapse to bound clumps}, thus developing non-axisymmetric features in the disk. Star formation within these clumps induces strong stellar feedback, and torques due to the non-axisymmetric features drive angular momentum outward and thus mass inward, which flows down the potential well towards a central bulge. Both stellar feedback and radial mass transport are sources of turbulence \citep{Krumholz18}, which raises the velocity dispersion and {consequently $Q$ above $Q_{\rm crit}$. When $Q$ grows above $Q_{\rm crit}$}, the disk stabilizes, clumps and non-axisymmetric features diminish, and turbulence-driving mechanisms weaken, allowing $\sigma$ to decrease and {$Q$ to eventually fall below $Q_{\rm crit}$ again}. {While other drivers of turbulence - both internal and external to the disk - such as magnetorotational instabilities, spiral shocks, and external accretion \citep{Federrath17,Federrath18,Ginzburg22} may play important roles at different masses and redshifts and should be incorporated into the picture in a self-consistent manner, this qualitative argument hints at a cyclical regulation mechanism that maintains the disk in a state of marginal stability, with {$Q\!\sim\!Q_{\rm crit}$}}. This was confirmed in numerical simulations of isolated disks {{\citep{Immeli04,BournaudElemegreen09,Hopkins12,Arora25}}}, as well as observations \citep{Genzel14,Fisher17}. However, caution should be taken when measuring $Q$ in observations, especially in clumps, because these measurements are usually dominated by the high surface density, non linear structures, where $Q$ is no longer meaningful.

In a cosmological setting, however, the situation is more complex. \citet{Inoue16} showed that, in the VELA simulations, clumps are formed out of `protoclump' regions, where the local, {two component $Q$ parameter, that is, the $Q$ parameter of a disk composed of stars and gas \citep{Rafikov2001,RomeoWiegert}}, is greater than {an order unity $Q_{\rm crit}$}, and occasionally substantially greater. This calls into question the validity of the local linear Toomre analysis in cosmological disks. Several other formation mechanisms for clumps are possible. For example, \citet{Inoue21} suggest a two stage formation of clumps, beginning with the development of spiral arms when $Q\!\sim\!2$, followed by the fragmentation of these arms to clumps when $Q\!\sim\!0.6$. However, this does not explain the clump formation in regions with $Q>3$ in VELA. Another possibility is the rapid dissipation of turbulence in perturbed regions. \citet{Elmegreen11} showed that if significant energy dissipation occurs within a crossing time, the threshold for stability increases to {$Q_{\rm crit}\sim\!2-3$}, again insufficient to explain the clump formation found by \citet{Inoue16}.  \citet{Lovelace78} derived a \textit{necessary} condition for non-axisymmetric perturbations to be unstable based on the existence of a local extremum of $f(r)=\Sigma\Omega/\kappa^2$. Some galaxies in the VELA simulations did show such a local extremum \citep{Inoue16}, so it is possible that some clumps are formed due to non-axisymmetric perturbations. Other theories \citep[e.g.][]{GrivGedalin12} eventually boil down to a stability criterion of {$Q_{\rm crit}\gtrsim2$}.

In \citet{Mandelker24} (henceforth \citetalias{Mandelker24}), we initiated the exploration of the possibility that clumps form out of protoclump regions with high levels of compressive turbulence. Broadly speaking, the turbulence velocity field can be thought of as composed of a compressive part, representing the local tendency to converge to or diverge from a point, and a solenoidal part, representing the local tendency to rotate about a point {\citep[see][for an extensive and rigorous overview]{Kritsuk07,Federrath10}}. {The former is represented by the divergence of the velocity field, while the latter by the curl of the velocity field (see Sect. \ref{sec:turbulence_decomposition}). While cosmological simulations, like the ones used in \citetalias{Mandelker24}, typically do not properly resolve the turbulence cascade, very high resolution isolated simulations do show proper cascade of such motions \citep{Bournaud10,Fensch23}.} In a steady state, isotropic and homogeneous turbulent field in equipartition, the global ratio of energy in compressive modes to total turbulent energy is $\sim\!1/3$. However, if for any reason there is a deviation from this value in favor of compressive modes, the turbulence itself is able to generate more dense regions \citep{Padoan02,Federrath10}, that can be sufficient for self gravity to take over, while having large velocity dispersions that keep $Q$ above unity. Indeed, \citet{Hopkins13b} found that, statistically, $Q$ parameters of $\gtrsim 5$\footnote{\citet{Hopkins13b} found that for highly supersonic turbulence, a disk is never \textit{statistically} stable. Their study focused on protoplanetary disks. However, adopting model parameters suited to galactic disks, where turbulence is driven by multiple mechanisms with significant morphological variations, suggests a more appropriate $Q$ threshold of $\sim\!5$.} are marginally stable in highly supersonic disks.

While inspecting both cosmological and isolated simulations, \citetalias{Mandelker24} found that protoclump regions in cosmological simulations do show an excess of energy in compressive modes, while protoclump regions in isolated simulations do not. It thus implies that the cosmological environment drives more compressive turbulence by physical processes that are absent an in isolated setting. In this work, we examine two possible causes for the excess of compressive modes of turbulence in a cosmological environment: compressive gravitational tides {\citep[a scenario in which $Q$ may be altered;][]{Jog14}} and stream-disk interaction. For the former, several studies showed that during galaxy mergers of other strong interactions, elevated levels of compressive tides induces clump and cluster formations, as well as high levels of compressive turbulence \citep{Renaud09,Renaud13,Renaud22,Li22}. Mergers can also induce clump formation by the wake they produce during the merger \citep{Nakazato}. For the latter, stream-disk interaction has the potential to drive turbulence due to the collisions between the flows (see discussion in Sect. \ref{sec:discussion_efficiency}).

The paper is organized as follows. In Sect. \ref{sec:methods} we introduce the methods used in the analysis, including the turbulence decomposition (Sect. \ref{sec:turbulence_decomposition}), the tidal field analysis (Sect. \ref{sec:tidal_field}) and the stream-disk interaction analysis (Sect. \ref{sec:stream_disk_interaction}). In Sect. \ref{sec:results} we present our main analysis - in Sect. \ref{sec:pc_vs_random} we present a comparison the fraction of energy in compressive modes, the tides and the presence of streams between protoclumps and random patches, and in Sect. \ref{sec:ftides_and_fstr_vs_fcomp} we correlate the different quantities. In Sect. \ref{sec:discussion} we discuss some caveats, and in Sect. \ref{sec:conclusions} we present our conclusions.
\section{Methods}\label{sec:methods}
\subsection{The VELA cosmological simulations}
We analyze eight galaxies from the {VELA3} suite of cosmological simulations\footnote{The galaxies we analyze are labeled V07, V08, V11, V12, V14, V25, V26 and V27, in accordance with table 1 in \citet{Mandelker17}.} \citep{Ceverino14,Mandelker17}. The simulations use the {ART} code \citep{Kravtsov97,Ceverino09}, which is a gravito-hydrodynamics, grid-based, adaptive mesh refinement code, with a maximal resolution of $\sim17-35$ proper $\pc$ at all times, down to $z\!\sim\!1$. The dark matter mass resolution is $8.3\!\cdot\! 10^4\Msun$, and the minimal stellar particle mass is $10^3\Msun$. The code includes gas and metal cooling, UV-background photoionization and self shielding in dense gas, stochastic star-formation, thermal feedback, radiation pressure and metal enrichment from stellar feedback \citep{Ceverino10,Ceverino12,Ceverino14}. A more detailed description of the simulations can be found in \citet{Ceverino14} and \citet{Mandelker17}. We use the high temporal resolution version of the simulations, which has $\sim30$ snapshots per disc orbital time.

All of the eight galaxies undergo a phase of compaction to a blue nugget \citep[sometimes multiple;][]{Zolotov15,Lapiner23}, which involves a rapid increase in the central gas density, followed by an intense starburst and subsequent inside-out quenching. These compaction events are usually triggered by a major merger \citep{Lapiner23}, although not exclusively. The galactic disk survives for long periods of time when the halo mass exceeds $\sim\!10^{11}\Msun$ \citep{Dekel2020}, which typically occurs after the compaction event. Thus, the main violent, disky\footnote{The galactic disk's dimensions and orientation are defined by those of the cylinder that contains $85\%$ of the cold component, consisting of gas with $T<10^4\,{\rm K}$ and young stars with ${\rm age}<100\,{\rm Myr}$, within $0.15{\rm R_{vir}}$ \citep{Mandelker17}.} phase of the galaxies typically lasts from $z\sim\!4-\!1$ \citep{Mandelker17,DekelMigration}, which is roughly the period of time analyzed in \citet{Inoue16} and \citetalias{Mandelker24}.
\subsection{Clump finder \& protoclumps}
We use the clump finder developed by \citet{Mandelker17}. In short, the 3D density field of baryons is dumped onto a uniform grid with grid spacing of $\Delta x=70\pc$. The grid is then smoothed with a spherical Gaussian with full-width-at-half-maximum of $W=2.5\kpc$. The residual, $\delta$, is then defined as $\delta = (\rho-\rho_W)/\rho$, where $\rho$ and $\rho_W$ are the raw and smoothed density fields, respectively. Clumps are defined as connected regions containing at least eight uni-grid cells with $\delta>10$, as defined using either stellar density, or the combined density of cold gas ($T<1.5\cdot10^4\ {\rm K}$) and young stars (stellar age $<100\ {\rm Myr}$).

Since VELA uses ART, which is a grid-based code, and the current run did not include tracer particles, clumps are tracked through time based on their stellar particles. We only track clumps that contain at least ten stellar particles. For each such clump at a given snapshot, we search for all clumps in the previous snapshot that contribute at least $25\%$ of their stellar particles to the current clump. If a given clump has more than one such progenitor, we consider the most massive one as the main progenitor, and the others as having merged with it. The formation time of the clump is considered to be the snapshot at which no progenitor was found in the two preceding snapshots. The clumps in VELA were studied in detail in \citet{Mandelker17}, and their properties are consistent with properties of observed clumps \citep{G18,Ginzburg21}.

To define protoclump regions, i.e. the regions out of which clumps form, we take the center-of-mass velocity of the clump at its initial formation snapshot in cylindrical coordinates ($v_r,v_\phi,v_z$), and extrapolate the position one snapshot back in time to determine the position of the protoclump. Assuming that the clump has just formed, and given the high temporal resolution, it is safe to assume that the clump's velocity is still attached to the overall disk velocity field. For the size of the protoclumps, since most of the collapsed clumps have radii in the range $R_{\rm C}\sim 100-300\pc$ \citep{Mandelker17}, and clumps usually contract by a factor of two to three \citep{Ceverino12,Dekel23}, we use a fixed radius of $R_{\rm PC}=0.5\,\kpc$ for all protoclumps. {We performed all of our analysis for $R_{\rm PC}=0.8\,\kpc$ with no qualitative difference}.
\subsection{Turbulence decomposition}\label{sec:turbulence_decomposition}
The heart of our analysis is based on compressive turbulence. Compressible turbulence in general, and compressive turbulence in particular\footnote{The former is the more general notion of allowing supersonic flows and shocks, while the latter generally refers to the tendency of the velocity field to converge to a point.} has been shown to impact the density distribution of interstellar gas \citep{Federrath10}, which in turn affects star formation in regions with excess of compressive turbulence. {Indeed, an extensive study by \citet{Federrath12} demonstrated, analytically and numerically, that the star formation rates, as well as the star formation efficiencies, increase by orders of magnitude when the turbulence driving is dominated by compressive modes.}

While compressive turbulence is conceptually simple to understand, it is mathematically more challenging to define. Below, we give two definitions, one local and one global, each having advantages but also drawbacks that present a caveat in our analysis. Discussion of these caveats is present in Sect. \ref{sec:discuss_definition_of_compressive_modes}.

A word on terminology - we call the mode of turbulence that describes a tendency to locally diverge or converge to a point the `compressive mode', and the mode of turbulence that describes a tendency to locally rotate about a point the `solenoidal mode'. Out of the two possible directions of the compressive mode, we have the `converging mode', for which $\nabla\cdot\vec{v}<0$, and the `diverging mode', for which $\nabla\cdot\vec{v}>0$.

\subsubsection{Global decomposition}\label{sec:global_decomposition}
Every vector field $\vec{v}(\vec{r})$ can be decomposed, using the Helmholtz decomposition, to curl-free, divergence-free and harmonic components
\begin{equation}
\vec{v}(\vec{r}) = \nabla\phi + \nabla\times\vec{A} + \nabla\psi   \equiv \vec{v}_{\rm comp} + \vec{v}_{\rm sol} + \vec{v}_{\rm har},
\end{equation}
where $\phi,\psi$ are scalar functions, with $\psi$ satisfying $\nabla^2\psi = 0$ and $\vec{A}$ is a vector function. In order to perform this decomposition, appropriate boundary conditions are needed. Two types boundary conditions are commonly assumed, which make the decomposition straightforward - vanishing at infinity or periodic - both of which are not applicable in the local case. For the former, $\vec{v}\to0$ as $|\vec{r}|\to\infty$, then necessarily $\psi=const$ since $\psi$ solves Laplace's equation with $\nabla\psi = 0$ at the boundary\footnote{Suppose $\nabla\psi=0$ on the boundary. Let $D_0$ be some arbitrarily chosen point on the boundary. Then, if $D$ is any point on the boundary, $\psi(D) - \psi(D_0) = \int \nabla\psi\cdot d\vec{l}=0$, where the integration is along the boundary. This shows that $\psi(D)=\psi(D_0)$ for all points $D$ on the boundary. Thus, $\psi$ solves the Laplace equation with $\psi=const.$ on the boundary. From the uniqueness theorem for Laplace's equation, $\psi=const.$ in the entire volume.}. The value of $\psi$ is not physically important, only its gradient, which is zero. For the latter type of boundary condition, in a periodic box, $\psi=0$, since a harmonic function cannot be periodic in every direction, and if $\vec{v}$ is periodic, $\nabla\psi$ must be periodic, hence $\psi$ must be periodic.

In either of these cases, a periodic box or an infinite domain where $\vec{v}\to0$ as $|\vec{r}|\to\infty$, the decomposition can be performed using the Fourier transform. This is achieved by projecting the Fourier transform of the velocity field along the wavevector $\vec{k}$ and perpendicular to $\vec{k}$,
$$
    {G}_\phi(\vec{k}) = \frac{i\vec{k}\cdot\tilde{\vec{v}}}{|\vec{k}|^2},\ \ \ \vec{G}_A(\vec{k}) = \frac{i\vec{k}\times\tilde{\vec{v}}}{|\vec{k}|^2}.
$$
Here, $\tilde{\vec{v}}$ is the Fourier transform of $\vec{v}$. Using these two functions, we can write the Fourier transform of $\vec{v}$ as
\begin{equation}\label{eq:fourier_decomp}
\tilde{\vec{v}} = -i\vec{k}G_\phi + i\vec{k}\times\vec{G}_A.
\end{equation}
The desired curl-free and divergence-free parts are determined by the inverse Fourier transform of the above two terms. It is evident from this decomposition that $\nabla\cdot \vec{v}_{\rm comp} = \nabla\cdot\vec{v}$ and $\nabla\times\vec{v}_{\rm sol} = \nabla\times\vec{v}$.

From Parseval's theorem, one finds that $\vec{v}_{\rm comp}$ and $\vec{v}_{\rm sol}$ are orthogonal in the global sense, i.e.
\begin{equation}
    \int|\vec{v}|^2d^3\vec{r} = \int|\vec{v}_{\rm comp}|^2d^3\vec{r} + \int|\vec{v}_{\rm sol}|^2d^3\vec{r}, 
\end{equation}
but {\bf not} in a local sense (i.e., $\vec{v}_{\rm comp}\cdot\vec{v}_{\rm sol}\neq0$). This can easily be understood from the uncertainty principle - the decomposition is locally orthogonal in $\vec{k}$-space, and therefore only globally orthogonal in $\vec{r}$-space \footnote{We have that $|\vec{v}|^2 = |\vec{v}_{\rm comp}|^2 + |\vec{v}_{\rm sol}|^2 + 2\vec{v}_{\rm comp}\cdot\vec{v}_{\rm sol}$, where the last term is, in general, not zero.}. Another caveat with this decomposition is that it is possible to construct a velocity field such that at a given point,$\vec{r}_0$, $\vec{v}(\vec{r}_0)=0$, while $\vec{v}_{\rm comp}(\vec{r}_0),\ \vec{v}_{\rm sol}(\vec{r}_0)\neq 0$. Thus, assuming that $\left|\vec{v}_{\rm comp}\right|^2$ and $\left|\vec{v}_{\rm sol}\right|^2$ represent some local energy densities in a region, even if relatively isolated, can introduce significant errors.

Nevertheless, we can use this decomposition to robustly define the global fraction of energy in compressive modes of turbulence \citep[see also][]{Federrath10,BruntFederrath14}, that is
\begin{equation}
    f_{\rm comp,glob} = \frac{\int\left|v_{\rm comp}\right|^2d^3\vec{r}}{\int\left|v\right|^2d^3\vec{r}}.
    \label{eq:fcomp_def_global}
\end{equation}
While it can be analytically proven, {it is intuitive to understand that in a fully isotropic and homogeneous turbulence}\footnote{{That is, a velocity field whose two-point correlation function is a function of the magnitude of the separation only}} in equipartition, $f_{\rm comp,glob}=1/3$ - converging or diverging to or from a point is a one dimensional radial motion, while rotation (characterized by the solenoidal mode) about a point is two dimensional. Furthermore, for a fully isotropic and homogeneous turbulence in equipartition, the converging mode is half the total compressive power.
\subsubsection{Local decomposition}\label{sec:local_decomposition}
Given the inherit non-locality of the Helmholtz decomposition, a different method is required to characterize the compressiveness of the turbulence field on small, protoclump scales. The most common approach is based on the following. 

If $\vec{v}(\vec{r})$ is the velocity at the center of the clump, $\vec{r}$, then one can estimate the turbulence velocity field in the vicinity of this point is $\delta\vec{v}\equiv \vec{v}(\vec{r}+\delta\vec{r}) - \vec{v}(\vec{r})$. Since protoclumps are only mild overdensities, and tend to rotate with the disk, the variation of the velocity field within them can be interpreted as turbulence on scales smaller than the protoclump. This filtering approach is a common mathematical approximation for the turbulence field when one cannot resolve all the relevant scales \citep{Schmidt06,Garnier09,Aluie13,Schmidt14,Semenov24}{, and methods stemming from this formalism have been applied to analyze turbulence in the context of disk instability in various galaxy simulations \citep{Agertz09a,Agertz09b,Bournaud14,Inoue16,Goldbaum,Renaud21,Ejd}}. Technically, this approach captures random motions about the local mean motion. A more formal approach requires detailed power spectrum analysis to show this obeys proper energy cascade. It is quite challenging to capture properly the energy cascade in numerical simulations, even in the relatively high resolution in our simulations {{\citep[e.g.][]{Kritsuk11,Federrath21,Semenov24}}}, however detailed analyses show that such a filtering approach captures the interaction between scales properly \citep{Semenov24}. {We do note that very high resolution simulations of isolated galaxies do show that disk galaxies exhibit proper turbulence cascade \citep{Bournaud10,Fensch23} While not properly resolved here, the converging flows in our simulations are somewhere along the beginning of the inertial range.}

{To linear order, $\delta{\vec{v}}$ can be approximated as}
\begin{equation}\label{eq:local_decomp}
    {\delta\vec{v} = \vec{v}(\vec{r}+\delta\vec{r})-\vec{v}(\vec{r}) \approx \frac{1}{3}\lbrac{\vec{\nabla}\cdot\vec{v}}\delta\vec{r} + \frac{1}{2}\vec{\omega}\times\delta\vec{r} + \overleftrightarrow{S}\delta\vec{r}},
\end{equation}
where $\vec{\omega}=\vec{\nabla}\times\vec{v}$ is the vorticity, and $\overleftrightarrow{S}$ is a traceless symmetric tensor representing shearing motion. While the second and third terms in eq. \ref{eq:local_decomp} are orthogonal, the shear term is, in general, not orthogonal to either of them. This raises a complication in interpreting each component individually as a contributor to the total energy in the turbulence field.

{As mentioned above, the deviation of the velocity field from the local mean rotation is interpreted as the turbulence velocity field. However, $\divr{\vec{v}}$ and $\curl{\vec{v}}$ properly capture this turbulent behavior even if the mean rotation is not subtracted, as long as it varies on scales larger than the scale of the protoclump\footnote{{If the rotation velocity, $\vec{v}_{\phi}$, is varies slowly on the scales of the protoclump, then $\divr{\vec{v}} = \divr{\lbrac{\vec{v}_{\phi}+\Delta\vec{v}}}\approx \divr{\lbrac{\Delta\vec{v}}}$}}.}

Other works tend to neglect the shearing term \citep{Kida90,Kritsuk07,Renaud13}, and then the total turbulent kinetic energy in a sphere of radius $R$ is
\begin{equation}
    E_K\approx \frac{1}{9}\lbrac{\divr{\vec{v}}}^2\int r^4 d^3\vec{r} + \frac{1}{4}\left|\curl{\vec{v}}\right|^2\int r^4\sin^2\theta d^3\vec{r},
\end{equation}
where the factor of $\sin^2\theta$ comes from the cross product between $\vec{\omega}$ and $\vec{r}$. After integrating, we get
\begin{equation}
    E_K \approx \frac{4\pi}{9}\frac{R^5}{5}\lbrac{\divr{\vec{v}}}^2 + \frac{2\pi}{3}\left|\curl{\vec{v}}\right|^2\frac{R^5}{5}.
\end{equation}
This motivates defining $|\vec\nabla\cdot\vec{v}|^2$ as the local energy in compressive modes\footnote{The divergence part is smaller than the curl part by a factor of $2/3$. This amounts to a factor of $\sim\!0.8$ in the local fraction as defined in eq. \ref{eq:fcomp_def}, which doesn't affect the qualitative results.}, and $\left|\curl{\vec{v}}\right|^2$ as the local energy in solenoidal modes\footnote{Other possible reasonings behind this definition are either the correspondence with the Helmholtz decomposition, eq. \ref{eq:fcomp_global_and_local_relation}, or from the fact that the viscous dissipation rate can be decomposed to a solenoidal and compressive part, as shown in Appendix \ref{sec:viscous_dissipation_proof}. Both of these explanations rely on specific boundary conditions.}. We take the same approach, and discuss its caveats and potential future directions in Sect. \ref{sec:discuss_definition_of_compressive_modes}.  We thus define the local fraction of energy in converging mode, at a given point, as
\begin{equation}
     f_{\rm conv,loc} = \frac{\left|\vec{\nabla}\cdot\vec{v}\right|^2_{\rm neg}}{\left|\vec{\nabla}\cdot\vec{v}\right|^2+\left|\vec{\nabla}\times\vec{v}\right|^2}
    ,
\end{equation}
where $\left|\vec{\nabla}\cdot\vec{v}\right|^2_{\rm neg}$ refers to regions with negative divergence, representing compression (as opposed to expansion). {As discussed in the Introduction, and shown by turbulence box simulations \citep{Federrath10,Semenov24}, an increasingly negative divergence lead to larger turbulence velocity dispersion, and in particular compressive turbulence, which promotes dense regions that can become self-gravitating}. For a given region, the total fraction of energy in converging modes is thus
\begin{equation}
    \fcomp = \frac{\int\left|\vec{\nabla}\cdot\vec{v}\right|^2_{\rm neg}d^3\vec{r}}{\int\left|\vec{\nabla}\cdot\vec{v}\right|^2d^3\vec{r}+\int\left|\vec{\nabla}\times\vec{v}\right|^2d^3\vec{r}}
    \label{eq:fcomp_def}.
\end{equation}
In Appendix \ref{app:global_vs_loca_proof}, we show that for a turbulent field in which the power spectrum of $\vec{v}_{\rm comp}$ and $\vec{v}_{\rm sol}$ are proportional, with the appropriate boundary conditions, when integrating over the entire volume,
\begin{equation} \label{eq:fcomp_global_and_local_relation}
    \frac{\int\left|v_{\rm comp}\right|^2d^3\vec{r}}{\int\left|v\right|^2d^3\vec{r}} = \frac{\int\left|\nabla\cdot\vec{v}\right|^2d^3\vec{r}}{\int\left|\vec{\nabla}\cdot\vec{v}\right|^2d^3\vec{r}+\int\left|\vec{\nabla}\times\vec{v}\right|^2d^3\vec{r}}.
\end{equation}
Thus, the definitions of eq. \ref{eq:fcomp_def} and \ref{eq:fcomp_def_global} agree, contingent on the appropriate boundary conditions.

To calculate the local fraction in our simulations, we follow the method of \citetalias{Mandelker24}. First, we dump our AMR grid onto a uniform grid with $0.2\ \kpc$ resolution. Then, we compute the nine derivatives of the velocity field, $\partial v_i/\partial r_j$, using a second order, centered finite differences method. Then, for a given region, we integrate over its volume the quantities $\left|\vec\nabla\cdot\vec v\right|_{\rm neg}^2$ (taking into account only cells with negative divergence), $\left|\vec\nabla\cdot\vec v\right|^2$ and $\left|\vec\nabla\times\vec v\right|^2$. We then define $\fcomp$ of a given
region, according to equation \ref{eq:fcomp_def}. As discussed in Sect. \ref{sec:global_decomposition}, a value of $1/3$ for the total power in compressive turbulence (compression or expansion) is expected in fully isotropic and homogeneous turbulence in equipartition, and thus a value of $1/6$ is expected for $\fcomp$. {While the turbulence in galactic disk is not necessarily isotropic and homogeneous (certainly not on scales larger than the scale height, where the turbulence becomes two-dimensional), we use these values as a reference, saying that regions with $\fcomp > 0.33$ have excess of converging modes.}

{We note that after the protoclump has collapsed to a clump it is expected to become rotation-supported, as shown in \citep{Ceverino12}, and thus $\fstr$ will go down.}

\subsection{Tidal field}\label{sec:tidal_field}
When an extended body resides in a gravitational field, different mass elements experience slightly different gravitational forces. In the rest frame of the body, this difference translates to compression or expansion along different directions. The tidal field is in general not isotropic, and can be compressive or expansive in different directions. In principle, it can be fully compressive (i.e. compressive in every direction), but not fully expansive (see below).

Fully compressive tides can arise in systems which reside in cosmological environments \citep{Renaud09}, but can also arise internally in isolated systems, even in smooth density profiles \citep[][]{Dekel03}. If a patch of the disk resides in a region where the tidal field is fully compressive, the turbulence in this region can become dominated by compression \citep{Renaud13}, and this region may undergo a gravitational collapse due to the tides, eventually forming a giant clump. Mathematically, the tidal field is quantified by the tidal tensor, to be explained next.

The first order approximation for the gravitational acceleration about a point $\vec{r}_0$ is
\begin{equation}
    F_j(\vec{r}_0+\delta\vec{r}) = F_j(\vec{r}_0) + \frac{\partial F_j}{\partial r_i}\delta r_i + \ldots,
\end{equation}
where we have employed the summation notation. If $\phi(\vec{r})$ is the gravitational potential, then $F_j = -\partial\phi/\partial r_j$, and we can rewrite the acceleration, to first order, as
\begin{equation}
    F_j(\vec{r}_0+\delta \vec{r}) \approx -\left.\frac{\partial\phi}{\partial r_j}\right|_{\vec{r}_0} - \left.\frac{\partial^2\phi}{\partial r_i\partial r_j}\right|_{\vec{r}_0}\delta r_i.
\end{equation}
We define the tidal tensor as $T_{ij} = \partial^2\phi/\partial r_i\partial r_j$. Under this definition, the first order approximation for the gravitational acceleration is:
\begin{equation}
    F_j(\vec{r}_0+\delta\vec{r}) \approx F_j(\vec{r}_0) -T_{ij}\delta r_i.
\end{equation}
The tidal tensor is a symmetric tensor, and {therefore its eigenvectors are orthogonal, i.e. we can find an orthogonal basis that diagonalizes $T_{ij}$}. The eigenvalues are real, and we order them by value $\lambda_1\geq\lambda_2\geq\lambda_3$. Using our definition, a positive eigenvalue represents compression along the direction of its corresponding eigenvector, while a negative value represents expansion along this direction\footnote{Note that other publications  \citep[e.g.][]{Renaud09} defined the tidal tensor with a minus sign. Under such definition, a positive (negative) eigenvalue represents disruption (compression) rather than compression (disruption).}. If all eigenvalues are positive, the tidal field is said to be fully compressive. Since $\lambda_1+\lambda_2+\lambda_3 = {\rm trace}(T) = \nabla^2\phi=4\pi G\rho\geq 0$, $\lambda_1$ is necessarily positive (i.e. the tidal field cannot be fully expansive), while $\lambda_2,\lambda_3$ can be either positive or negative. 

{Previous studies quantified the compressiveness of the tidal field solely by the third eigenvalue, $\lambda_3$ \citep{Renaud09,Li22}. However this approach is quite conservative, as while $\lambda_3$ might be negative, $\lambda_1$ and $\lambda_2$ can simultaneously be positive, meaning that the tidal compression along their corresponding directions can be substantial. We therefore take a different approach, and define a quantity, $\ftides$, which takes into account all eigenvalues.} Assume a spherical region of size $R$ and constant density. As stated above, the tidal tensor is symmetric, and therefore we can align the coordinate system with its three eigenvectors. The mean value of the acceleration along the radial direction (indicative of compression/expansion) is then
\begin{equation*}
\begin{split}
    F_t &=-\frac{1}{4\pi/3 R^3} \int \lbrac{\overleftrightarrow{T}\vec{r}}\cdot\frac{\vec{r}}{r}d^3\vec{r} =-\frac{1}{4\pi/3 R^3}\int T_{ij} r_j\frac{r_i}{r}d^3\vec{r}\\
    &= -\frac{1}{4\pi/3 R^3}\int\frac{\lambda_1 x^2+\lambda_2y^2+\lambda_3z^2}{r}d^3\vec{r} = -\frac{1}{4}R\lbrac{\lambda_1+\lambda_2+\lambda_3} \\
    &\equiv -\frac{1}{4}R\lambda_1\lbrac{1+f_{\rm tides}},
\end{split}
\end{equation*}
where we have defined
\begin{equation}
\label{eq:f_tides}
    f_{\rm tides}=\frac{\lambda_2+\lambda_3}{\lambda_1}.
\end{equation}
\begin{figure}
    \centering
    \includegraphics[scale=0.5]{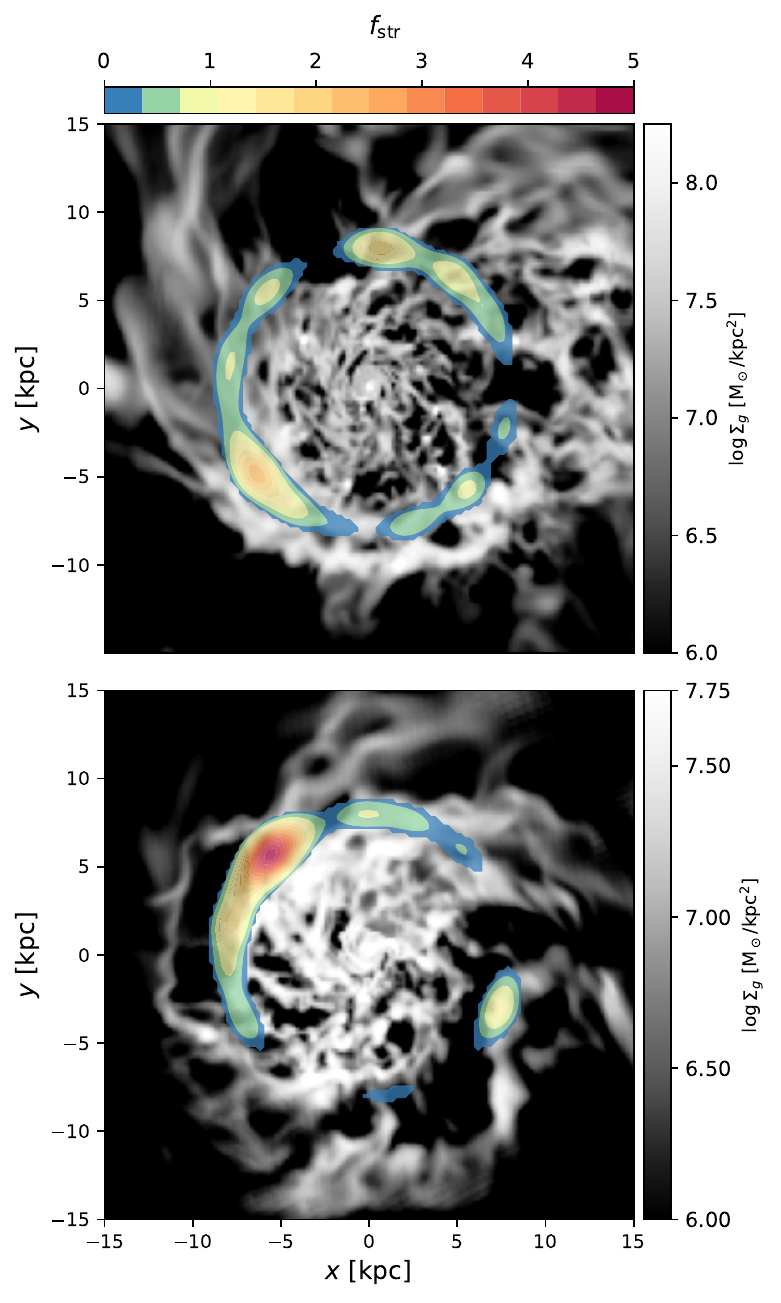}
    \caption{Proof of concept for $\fstreams$. The background color shows the projected surface density of gas in ${\rm V07}$ at $z\sim\!2$ (top) and in ${\rm V08}$ at $z\sim\!1$ (bottom). The contours are of $\fstreams$ in angular bins at a radius of $8\ \kpc$ in both panels. We can see how $\fstr$ increases in regions where the stream interacts with the disk.}
    \label{fig:fstr_example}
\end{figure}
This simple calculation motivates us to use $\ftides$ as a quantification of how compressive or expansive the tidal field is in a given spherical region - if $\ftides>0$, then the tidal field is substantially compressive along at least two directions (those of $\lambda_1$ and $\lambda_2$), and potentially fully compressive. If $\ftides < 0$, at least along one direction (that of $\lambda_3$), and potentially two directions (those of $\lambda_2$ and $\lambda_3$), the tidal field is substantially expansive\footnote{Our calculation resulted in {$F_t = -(R/4)\lambda_1(1+\ftides)$}, which is positive only when $\ftides<-1$, meaning that the {\bf mean} acceleration is positive only when $\ftides<-1$. Nevertheless, substantial tidal stripping can occur along a particular direction even if the mean acceleration is negative.}. As an example, for a spherically symmetric density profile, $\ftides = 2-\alpha(r)$, where $\alpha(r)$ is the logarithmic slope of the average density within $r$ \citep{Dekel03}. For values of $\alpha=0,1,2,3$, corresponding to a flat core, a cuspy profile, an isothermal sphere and a point mass, respectively, $\ftides = 2,1,0,-1$. Indeed, for a cored profile and a cuspy profile, the average tidal field is compressive. For a cored profile it is fully compressive in all three directions, while for a cuspy profile it is zero along the radial direction and compressive along the other two directions. For an isothermal sphere, the tidal field along the radial direction is expansive and equal in magnitude to the force along the other two directions, which are compressive. The average tidal force is thus zero. Finally, for a point mass, the tidal force along the radial direction is the strongest, and is expansive. {After the protoclump collapses, it will dominate the potential in its vicinity, thus $\ftides$ loses its ability to quantify tides induced by the larger environment of the galaxy.}

To calculate the compressiveness of the tidal field in a particular region in our simulations, we first dump the gravitational potential onto the same grid uniform grid as in Sect. \ref{sec:local_decomposition}, and calculate the Hessian matrix using a second order, central finite differences method. {Given that the resolution of the simulation in the regions of interest is $\sim\!20\!-\!100\,\pc$, a grid of resolution of $0.2\,\kpc$ is sufficient to be well above the effective gravitational softening induced by the Poisson solver. We did perform similar analysis on a $0.4\,\kpc$ resolution grid, and found no qualitative differences.} We then diagonalize the matrix at every grid cell (using the known analytical equations for $3\!\times\!3$ matrices) to get the three eigenvalues for each cell. Finally, we perform a volume-weighted average of each eigenvalue inside the given region, and plug the values into the definition of eq. \ref{eq:f_tides}, to get $\ftides$ for each region of interest.

\begin{figure*}
    \centering
    \includegraphics[width=\hsize]{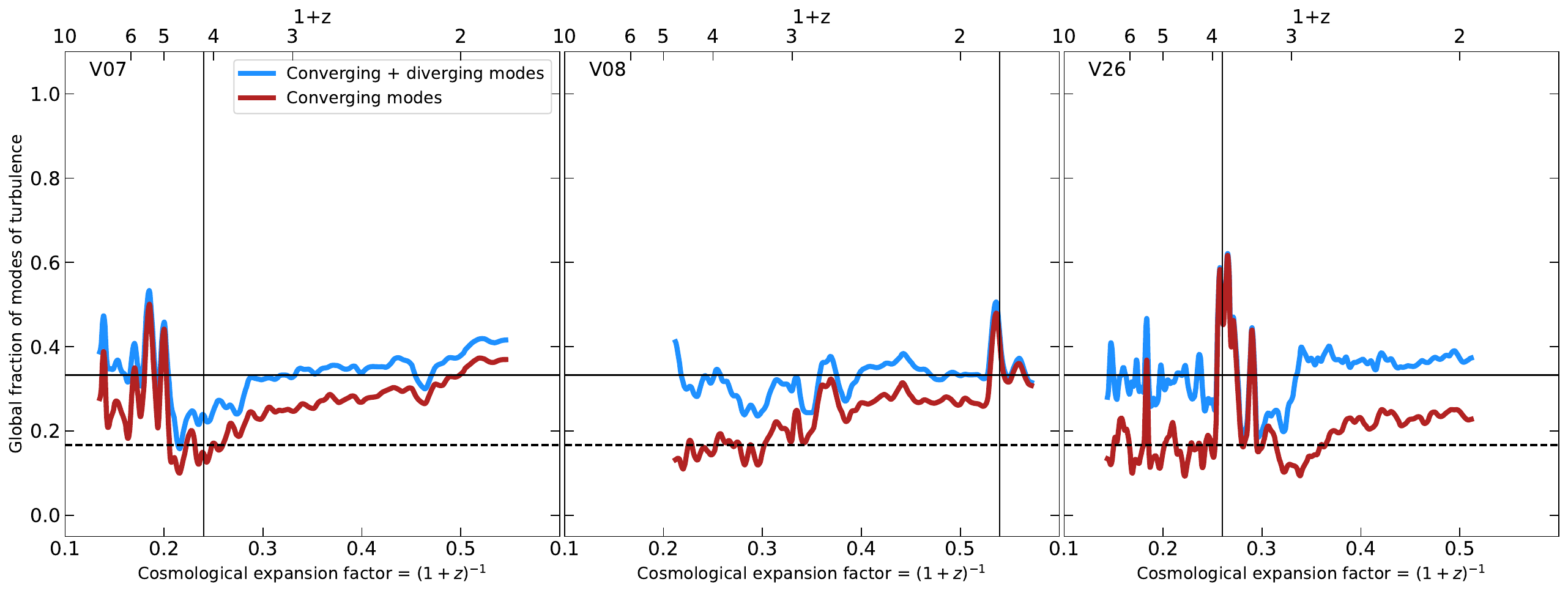}
    \caption{The temporal evolution of the global fraction of turbulent energy in converging and diverging (blue curve) and only converging (i.e. only cells with $\vec\nabla\cdot\vec v<0$; red curve) flows, for three galaxies from our suite. The vertical black line indicates the moment of the blue nugget phase in each galaxy \citep{Lapiner23}. The horizontal solid (dashed) lines indicate one third (sixth) of the total energy, values expected for fully isotropic and homogeneous turbulence in equipartition. Each of the blue nugget phases are preceded by at least one major merger, which increases the global fraction in compressive turbulence, before settling back to the equilibrium values.}
    \label{fig:global_fcomp_vs_time}
\end{figure*}
\begin{figure}
    \centering
    \includegraphics[width=\linewidth]{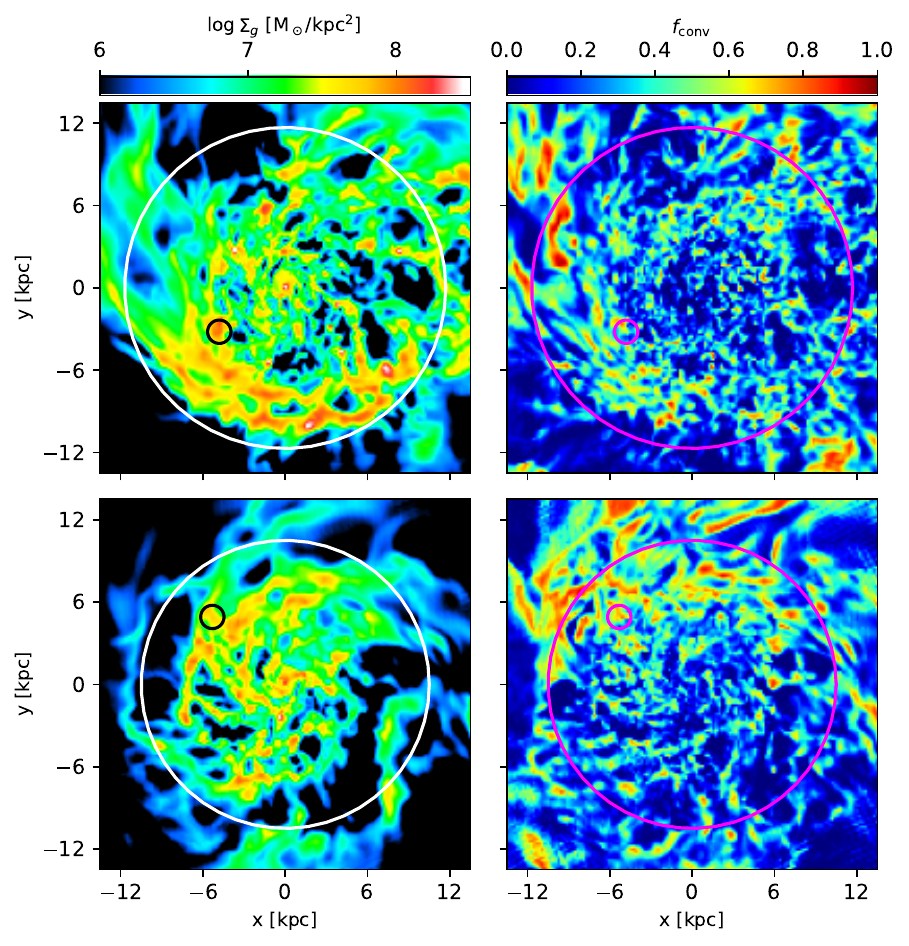}
    \caption{Correlation of $\fcomp$ with gas density in protoclump regions. In the left panels, we show the projected surface density of the gas. The top row represents V07 at $z\sim 2$ while the bottom row represents V08 at $z\sim 1$, as in figure \ref{fig:fstr_example}. The black circles indicate the location of a protoclump region. The white circle is the disk radius, defined as the radius that contains $85\%$ of the cold gas in the disk. In the right panels, we show maps of {mass weighted projections of} $\fcomp$ in the same snapshot. The small magenta circle shows the location of the same protoclump region, while the large magenta circle is the disk radius. We can see that inside the region of the protoclump, most of the mass has $\fcomp\sim0.5-0.8$. Furthermore, we see that the protoclump regions are close to where the stream joins the disk (see Sect. \ref{sec:stream_disk_interaction} and figure \ref{fig:fstr_example}).}
    \label{fig:fcomp_examples}
\end{figure}
\subsection{Stream-disk interaction}\label{sec:stream_disk_interaction}
When the streams interact with the disk, intense shocks are expected to occur {due to the collision of the stream material with the disk material,} and turbulence is expected to be stirred \citep{Ginzburg22}, especially if the streams are dense and clumpy due to fragmentation \citep{KlessenHennebelle,Forbes23}. Indeed, an ongoing series of papers, studying the evolution of cold streams feeding massive halos, has revealed that streams are expected to fragment into dense clumps, either gravitationally \citep{Mandelker18,Aung19}, or due to cooling \citep{Mandelker20a,Aung24}. Furthermore, since the streams flow towards the central galaxy in the potential of the dark matter halo, gravitational focusing tends to make the streams narrower and denser as they reach the disk \citep{Aung24}. Thus, we expect regions of the disk that interact with streams to be sites of high compressive turbulence that can lead to clump formation. This was briefly discussed in \citetalias{Mandelker24}, and can be seen in figure 3 therein. The degree to which stream-disk interaction drives turbulence is a subject of an ongoing work \citep[][Ginzburg et al., in prep]{Ginzburg22}.

In the current work, we want to understand whether or not protoclump regions are also sites of stream-disk interaction, and how the local fraction of compressive turbulence in these regions differs from other regions in the disk that do not interact with streams. The VELA simulations are AMR hydro simulations that do not have tracer particles. It is therefore difficult to determine whether a mass element was recently brought by accretion or has been part of the disk for a long time. We therefore use the approximate streamline method developed by \citet{DuttaChowdhury24}. In short, by assuming that the gas velocity field is roughly constant over a disk dynamical time, gas cells in a given snapshot can be traced using the gas velocity data of that snapshot only. Starting from its current 3D position, a gas cell in a given snapshot is traced back in time along its streamline to yield an approximate initial 3D position one dynamical time ago. If the distance between the initial and current positions is larger than $10\%$ the disk radius, we tag this cell as a stream material. We refer the reader to \citet{DuttaChowdhury24} for a more elaborate description of the method.

To quantify whether or not a given protoclump region resides in a site where a stream is interacting with the disk, we look at an annulus at the same radius of the protoclump, with its width being the size of the protoclump, namely $\Delta R = 2R_{\rm PC} = 1\,\kpc$. We then divide the annulus to angular bins of angular opening $\Delta\theta = d_{\rm PC}/r_{\rm PC}$, where $d_{\rm PC} = 1\ \kpc$ is the diameter of the protoclump region, and $r_{\rm PC}$ is the distance of the protoclump region from the galactic center. We then define
\begin{equation}\label{eq:fstr_def}
    \fstreams = \frac{M_{\rm stream,PC}}{\mean{M_{\rm stream,slice}}},
\end{equation}
where $M_{\rm stream,PC}$ is the total mass of stream material in the protoclump's angular bin, and $\mean{M_{\rm stream,slice}}$ is the average stream mass in all angular bins within the annulus. In figure \ref{fig:fstr_example} we show contours of $\fstreams$ at $8\,\kpc$ from $V07$ at $z\sim\!2$ and from $V08$ at $z\sim\!1$. We can see the stream from the top left joining the disk, resulting in an increased $\fstreams$ where the stream joins the disk.

{After the clump forms, it is either quickly disrupted by feedback or lives for a long time \citep{Mandelker17,Dekel23} and migrates to the center \citep{DekelMigration}. In the latter case, this will mean that $\fstr$ will go down as the clump migrates.}

\section{Results}\label{sec:results}
We want to test whether or not protoclump regions form in regions with fully compressive tides and/or in sites of stream-disk interaction. Furthermore, we would like to quantify the fraction of energy in compressive modes in such regions, and in converging modes in particular, and compare to the excess in converging modes found in protoclump regions in \citetalias{Mandelker24}. In order to differentiate protoclumps from the underlying disk, we associate to each protoclump a random patch that has the same distance from the galactic center and of the same size as the protoclump region, making sure it does not overlap with other protoclump regions. In the following analysis, we analyze protoclumps corresponding to clumps with a maximal baryonic mass larger than $10^8\ \Msun$, similar to \citetalias{Mandelker24}. We calculate $\fcomp,\ftides$ and $\fstreams$ as defined in eqs. \ref{eq:fcomp_def}, \ref{eq:f_tides} and \ref{eq:fstr_def} for the protoclump regions and the random patches in the disk, as outlined at the end of Sect. \ref{sec:local_decomposition}, Sect. \ref{sec:tidal_field} and Sect. \ref{sec:stream_disk_interaction}. 

In Sect. \ref{sec:global_fraction_of_compressive_turbulence} we analyze the global fraction in compressive modes as a function of time. In Sect. \ref{sec:pc_vs_random} we compare the various quantities between protoclumps and their corresponding random patches, and in Sect. \ref{sec:ftides_and_fstr_vs_fcomp} we correlate the different quantities against the $\fcomp$.

\subsection{Global fraction of compressive turbulence}\label{sec:global_fraction_of_compressive_turbulence}
In figure \ref{fig:global_fcomp_vs_time}, we show the fraction of turbulent energy in converging or compressive (converging plus diverging) modes, averaged over the whole disk, as defined in eq. \ref{eq:fcomp_global_and_local_relation}, as a function of the cosmological scale factor. We show the evolution of three out of the eight galaxies we analyze. We can see that, most of the time, the total fraction of energy in the compressive mode is about $1/3$ (blue curve), as expected in fully isotropic and homogeneous turbulence in equipartition (see Sect. \ref{sec:global_decomposition}). The red curve shows only the fraction in converging modes, i.e. only for cells with negative divergence. We can see that for V26 it is close to $1/6$, while for V07 it is usually around $0.3$. For V08, at early times, the total fraction of energy in converging modes is around $1/6$, increasing to $1/3$ at later times.

Each of these galaxies undergoes a compaction event that leads to a blue nugget \citep{Zolotov15}, which is usually preceded by a major merger \citep{Lapiner23}. We can see that before the major mergers and the subsequent compaction events occur, the fraction of energy in compressive modes (converging plus diverging) is roughly $1/3$, as expected in a fully isotropic and homoegeneous turbulence in equipartition\footnote{As can be seen in figure 1 in \citet{Renaud13}, a galaxy in isolation attains equipartition within a couple of dynamical times.}. During the merger, we can see a sharp increase in the fraction of energy in global compression, similar behavior as seen in idealized simulations of galactic mergers \citep{Renaud13}. {As we show below in Sect. \ref{sec:ftides_results} and figure \ref{fig:ftides_hist}, we find that clump formation is indeed correlated with compressive tides, which are expected to arise during mergers.}

Overall, we strengthen the result of \citetalias{Mandelker24} that globally, the turbulence field does not deviate much from the equilibrium ratios, except when undergoing major mergers and compaction events.

\begin{figure}
    \centering
    \includegraphics[width=\linewidth]{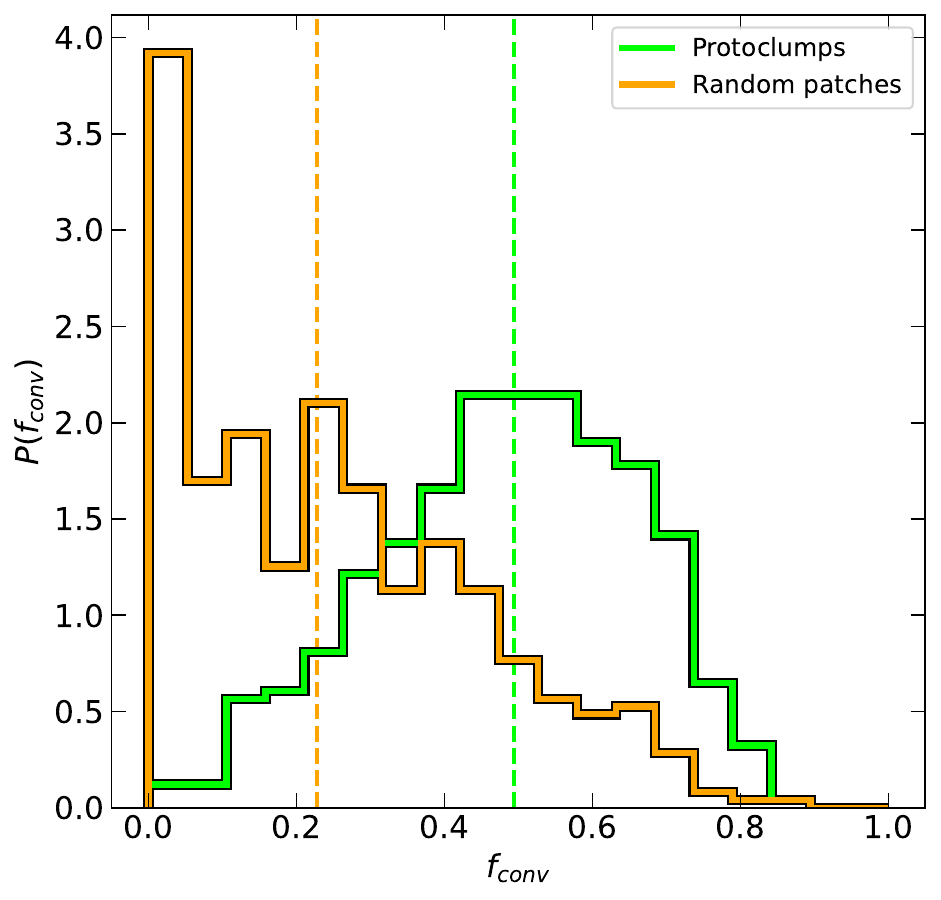}
    \caption{The probability distribution of $\fcomp$ in protoclump regions (green histogram) and random patches (orange histogram) over all eight galaxies. The vertical lines of each color indicate the median of the corresponding distribution. We can see that protoclump regions have a median $\fcomp\sim 0.5$, while random patches have a median $\fcomp\sim 0.21$, meaning that protoclump regions have a strong excess in converging modes, compared to random patches in the disk, which are more inline with equilibrium values.}
    \label{fig:fcomp_hist}
\end{figure}

\subsection{Protoclumps vs. random patches}\label{sec:pc_vs_random}
We now turn to the local analysis in small regions in the disk. We start by comparing $\fcomp, \ftides$ and $\fstreams$ between protoclump regions and random patches (as explained in the beginning of this section).
\begin{figure*}
    \centering
    \includegraphics[width=0.45\linewidth]{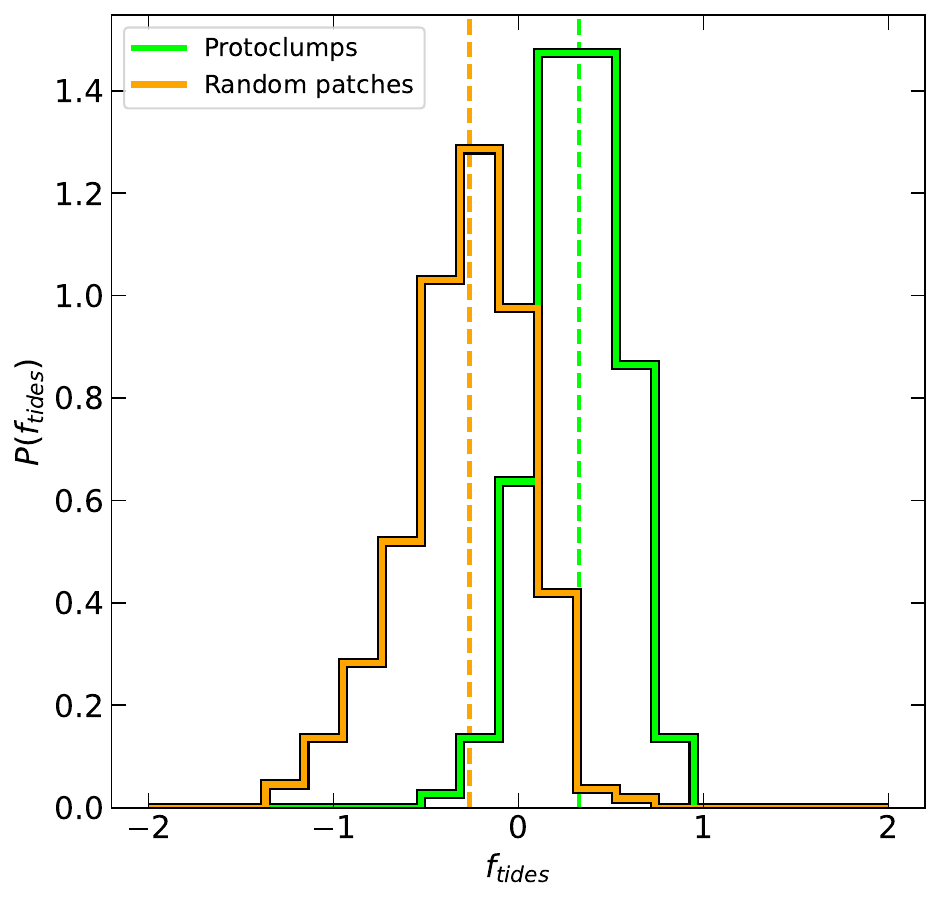}\smallskip
    \includegraphics[width=0.45\linewidth]{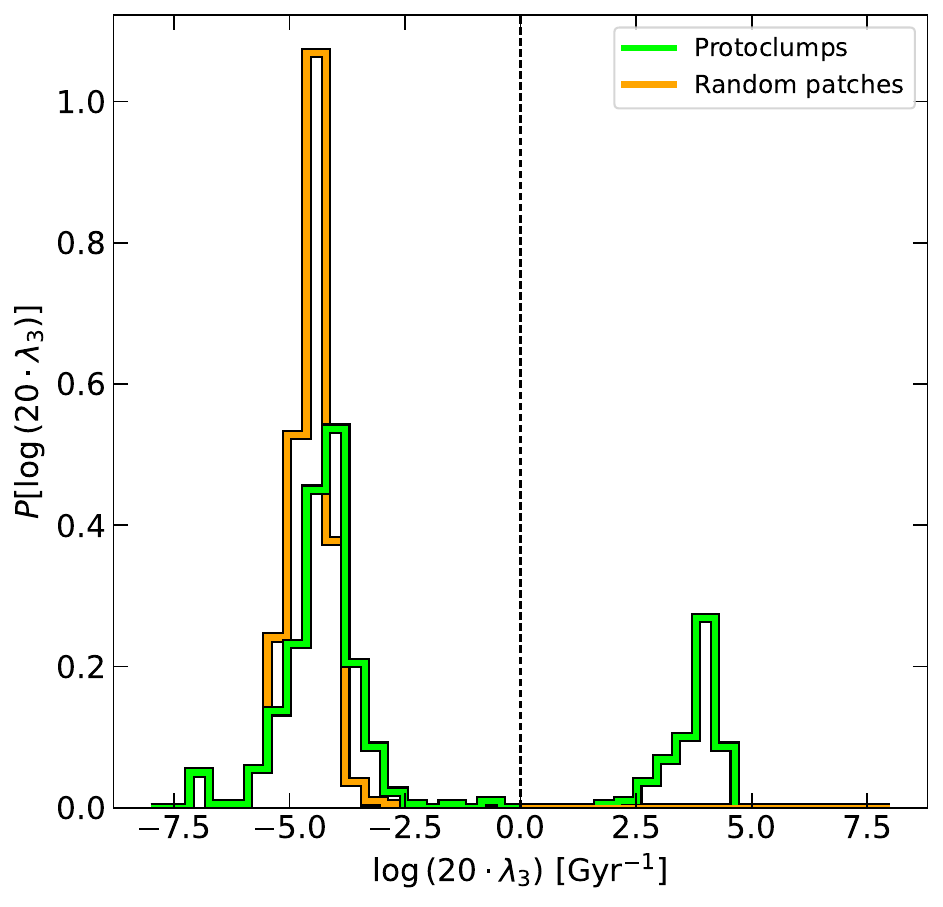}
    \caption{Probability distributions of $\ftides$ (left) and $\lambda_3$ (right) in protoclump regions (green histograms) and random patches (orange histograms). \textit{Left}: The vertical lines of each color indicate the median of the corresponding distribution, $0.33$ and $-0.27$ for protoclumps and random patches, respectively. $\ftides$ in protoclump regions is mostly positive, indicating substantial or fully compressive tides, while random patches mostly have $\ftides<0$, indicating substantial stripping. \textit{Right}: $\lambda_3$ was scaled by $20$ in order to separate negative and positive values and present them in a logarithmic scale. Negative values are actually $-\log(-20\cdot\lambda_3)$ (e.g. a value of $-5$ along the x-axis represents $-10^{5}\ {\rm Gyr^{-1}}$). $\lambda_3$ is positive almost exclusively in protoclump regions, and about $\sim25\%$ of the protoclump regions have $\lambda_3>0$, implying fully compressive tides.}
    \label{fig:ftides_hist}
\end{figure*}
\subsubsection{Compressive turbulence}\label{sec:fcomp_results}
In figure \ref{fig:fcomp_examples} we show examples of snapshots from ${\rm V07}$ and ${\rm V08}$ at redshifts $z\!\sim\!2$ and $1$, respectively. The small black (magenta) circle in the left (right) panel marks a protoclump region. We can see that {the two protoclump regions presented in figure \ref{fig:fcomp_examples}} reside in a part of the disk that interacts with a stream that comes from the top left corner (see also figure \ref{fig:fstr_example}). As we have argued above, this interaction {can stir up turbulence, and can induce compression by local shocks which boosts the fraction of energy in converging modes}, as clearly seen in the right panel of figure \ref{fig:fcomp_examples}, which shows the mass weighted projection of $\fcomp$. {The same protoclump regions show elevated levels of converging modes of turbulence. Figure \ref{fig:fcomp_examples} shows other regions in the disk that have high values of $\fcomp$. While some of these may be other protoclumps, perhaps of clumps that were not identified by the clump finder, most of these regions are not dense enough to initiate gravitational collapse, or are susceptible to shear (as discussed in Sect. \ref{sec:discuss_definition_of_compressive_modes}).}

In figure \ref{fig:fcomp_hist}, we show the distribution of $\fcomp$ in all protoclump regions and their corresponding random patches, for all of the eight galaxies. We can clearly see that $\fcomp$ has very different distributions in protoclumps and random patches - protoclump regions have a median $\fcomp\sim 0.5$, larger than the expected equilibrium value, while random patches have a median of $\fcomp\!\sim0.21$, which is more in line with the expected equilibrium value of $\sim\!0.16$. These results strengthen our results from \citetalias{Mandelker24}, in which the analysis was performed only on $V07$.

\subsubsection{Tides}\label{sec:ftides_results}
In the left panel of figure \ref{fig:ftides_hist}, we show the distribution of $\ftides$ in protoclump regions and random patches. Recall from Sect. \ref{sec:tidal_field} that a positive value of $\ftides$ is indicative of a substantial or fully compressive tidal field, while a negative value hints at substantial stripping along some direction. Figure \ref{fig:ftides_hist} clearly shows that $\ftides$ has different distributions among protoclumps and random patches. While random patches have a median $\ftides\sim-0.26$, protoclump regions have a median $\ftides\sim0.32$, with the vast majority of protoclumps having $\ftides>0$.

In the right panel of figure \ref{fig:ftides_hist}, we show the distribution of $\lambda_3$ (arbitrarily scaled, see caption) in protoclump regions and random patches. Although the distribution is of $\log\lambda_3$, negative values on the x-axis represent negative values of $\lambda_3$ rather than values $0<\lambda_3<1\ {\rm Gyr^{-1}}$ (see caption). We can see that the tidal field in $\sim\!25\%$ of the protoclumps is fully compressive, since $\lambda_3>0$ and therefore all of the eigenvalues of the tidal tensor are positive. On the other hand all of the random patches have $\lambda_3<0$, meaning that the tidal field is expansive along at least one direction. {While this may hint that a consequence of clump formation is local, fully compressive gravitational tides, we remind the reader that the protoclump regions are mild, non-self gravitating \citep[with $Q\gg 1$;][]{Inoue16}, local overdensities. Thus, it is unlikely that the cause for the compressive tides is due to the clump formation process, but rather due to sources external to the protoclump. A detailed study of the source of gravitational tides is needed to make this distinction (see discussion in Sect. \ref{sec:discussion_source_of_tides}).}

We conclude from figure \ref{fig:ftides_hist} that almost all of the protoclumps reside in regions where the tidal field is substantially compressive, and almost exclusively, regions with fully compressive tides are protoclumps.

\begin{figure}
    \centering
    \includegraphics[width=\linewidth]{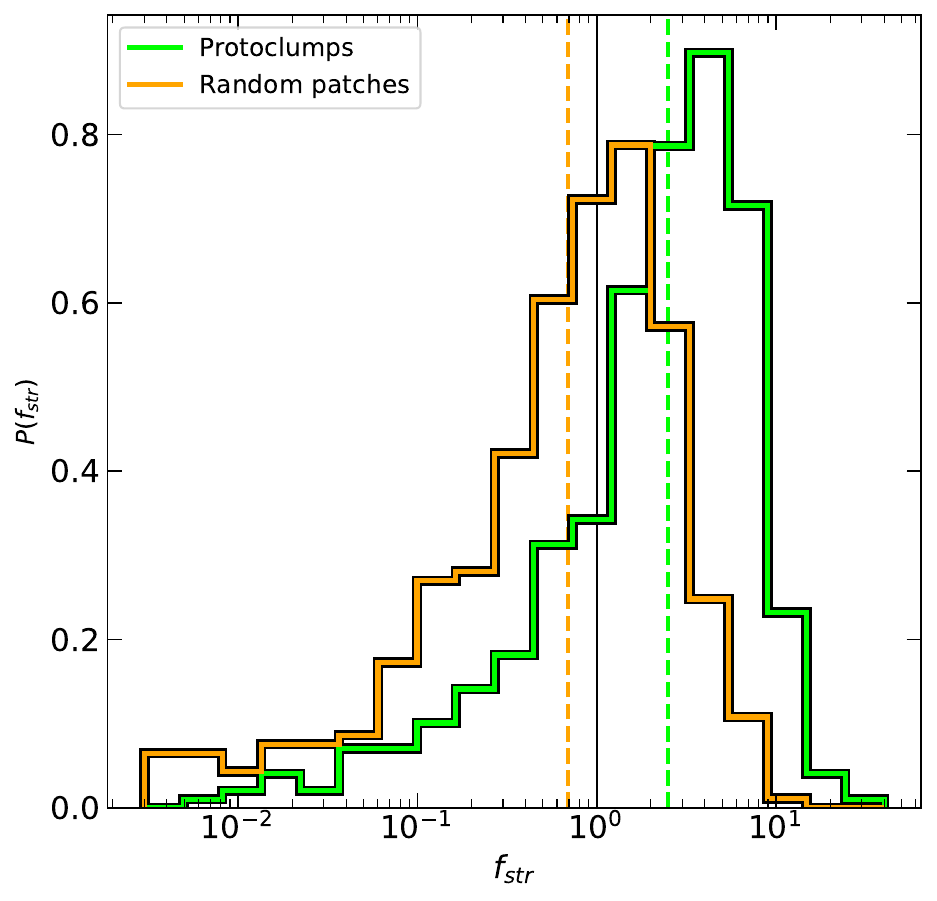}
    \caption{Same as figure \ref{fig:fcomp_hist}, but for $\fstr$, which measures the amount of stream mass in a given region (see text). The vertical black line is at $\fstr=1$. We can see that protoclump regions tend to have larger $\fstr$ than random patches, indicating that they reside in regions with intense accretion. The median $\fstr$ for protoclumps is $2.5$, while for random patches it is $0.8$.}
    \label{fig:fstr_hist}
\end{figure}
\subsubsection{Stream-disk interaction}\label{sec:fstr_results}
In figure \ref{fig:fstr_hist}, we show the distribution of $\fstr$ in protoclump regions and in their corresponding random patches. Random patches have, on average, $\fstr\sim\!0.8$, meaning that the stream mass in their vicinity is slightly less than the average over the annulus. For protoclump regions, we find on average $\fstr\sim\!2.5$, with a broad tail reaching $\fstr\!\sim\!10$, i.e. some protoclump regions have ten times more stream material than the average at their galactocentric radius, indicating that they reside in a site of stream-disk interaction.

The distributions of $\fstr$ in protoclumps and random patches are more similar than they are for both $\fcomp$ and $\ftides$, as can be seen by comparing figures \ref{fig:fcomp_hist} and \ref{fig:ftides_hist} (left panel) to figure \ref{fig:fstr_hist}. {However, recall that a major caveat in our defintion of $\fstr$ is the crude definition of stream material compared to disk material, due to the lack of tracer particles in our simulations \citep{DuttaChowdhury24}}. Furthermore, an energy-based or momentum-based quantity would perhaps be more appropriate than the mass-based quantity we use here, however without tracer particles these will suffer similar uncertainties. Nevertheless, we find that $\fstr=1$ is around the $30$-th percentile of the distribution for protoclump regions, meaning that around $70\%$ of the protoclumps reside in sites of stream-disk interaction, compared to only $\sim40\%$ of random patches.

\subsection{$\ftides$ and $\fstr$ vs. $\fcomp$} \label{sec:ftides_and_fstr_vs_fcomp}
In the previous section, we have found that protoclump regions tend to reside in regions where the turbulent velocity field is locally compressive (Sect. \ref{sec:fcomp_results}), tides are substantially or fully compressive (Sect. \ref{sec:ftides_results}), and are sites of stream-disk interaction (Sect. \ref{sec:fstr_results}). We now attempt to answer the question whether the tides or the stream-disk interactions cause the excess compression in the turbulent velocity field. In our current numerical setting it is not straightforward to draw causal conclusions on whether the tides or the stream-disk interaction are responsible for the excess of compressive turbulence. We therefore look at correlations between the quantities, leaving a more detailed physical analysis to future work (Ginzburg et al., in prep).

\begin{figure*}
    \includegraphics[width=0.5\linewidth]{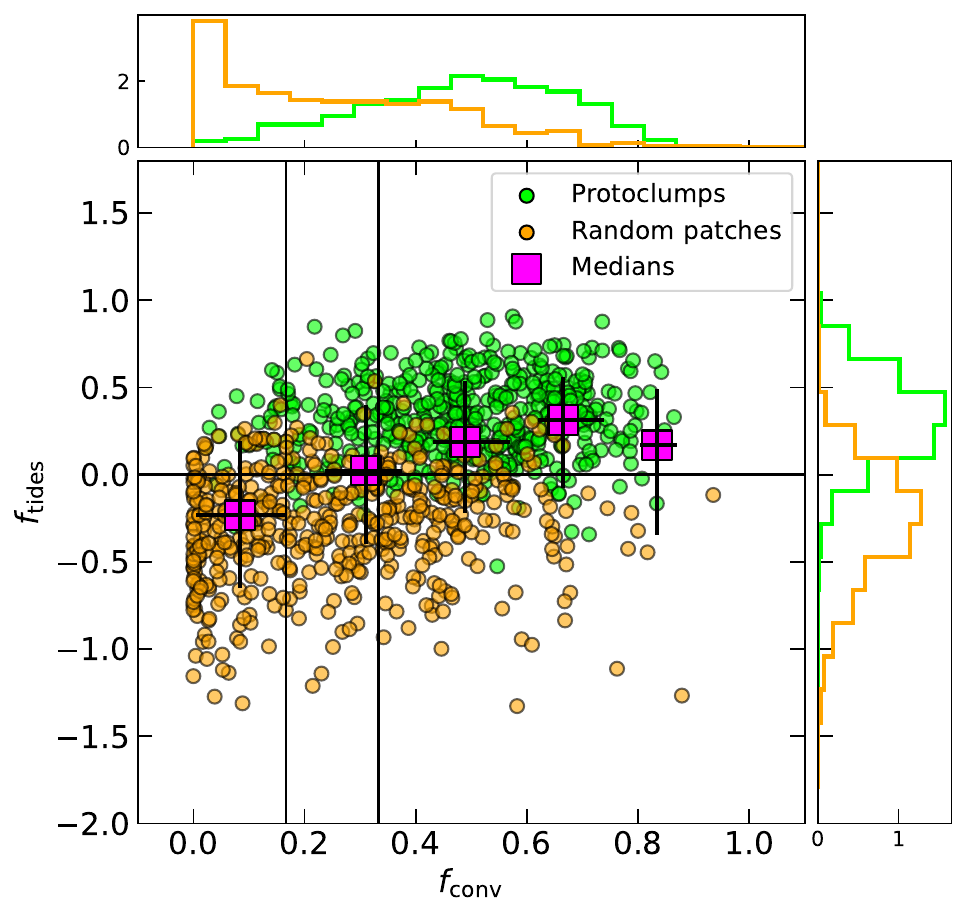}\smallskip
    \includegraphics[width=0.5\linewidth]{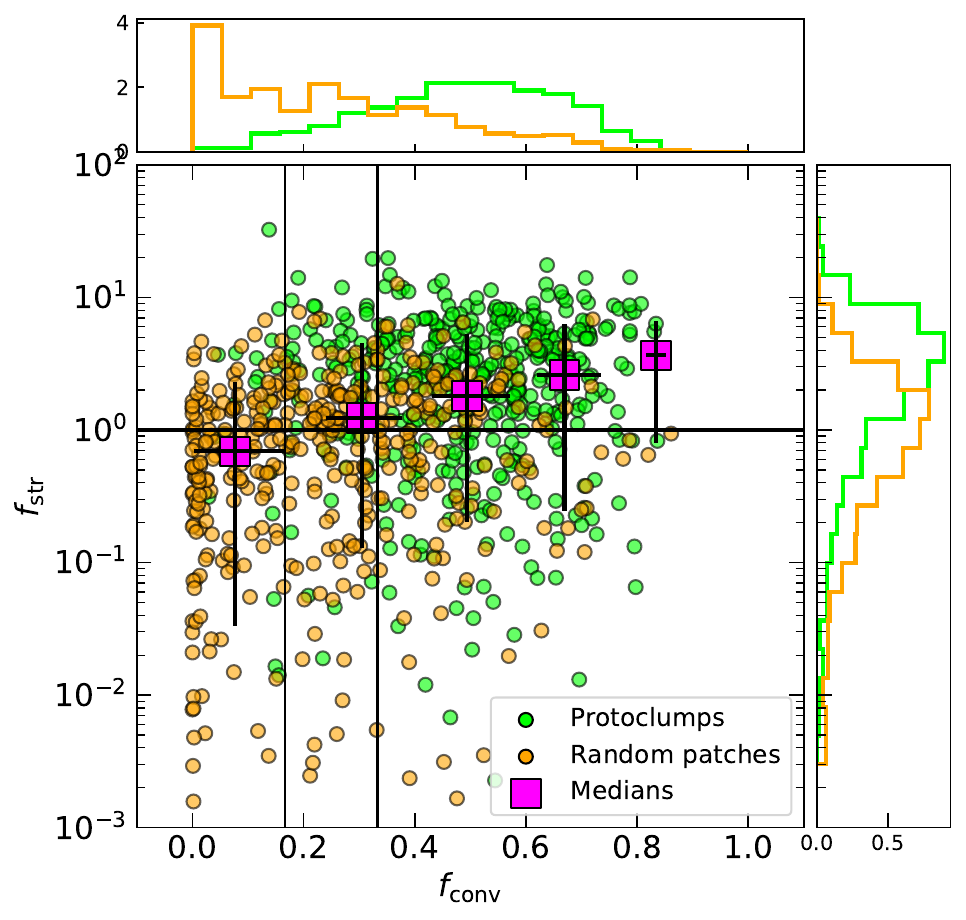}
    \caption{Correlations between $\fcomp$ (x-axis) and $\ftides$ (left) or $\fstr$ (right). Each point is either a protoclump (green points) or a random patch (orange points). The histograms are the projected distributions of the corresponding quantity. The vertical black lines correspond to $\fcomp=16\%$ and $\fcomp= 33\%$. The magenta squares are the medians of all of the points in bins of $\fcomp$, and the error bars are the $16$-th and $84$-th percentiles of each quantity. Both $\ftides$ and $\fstr$ show a positive correlation with $\fcomp$. The Spearman correlation coefficient for $\ftides$ and $\fstr$ against $\fcomp$ are $0.46$ and $0.3$, {with $p$-values of $2\cdot10^{-3},5\cdot10^{-3}$, respectively.} respectively. Furthermore, protoclump regions occupy the upper-right quadrants in both panels.}
    \label{fig:fcomp_vs_quantities}
\end{figure*}

The left panel of figure \ref{fig:fcomp_vs_quantities} shows $\ftides$ vs. $\fcomp$ for protoclumps and random patches. We can see that the median values of $\ftides$ in bins of $\fcomp$ increase with increasing $\fcomp$, from $\ftides\lesssim0$ at $\fcomp\sim0.1$ to $\ftides\sim0.5$ at $\fcomp\sim 0.65$. The Spearman correlation coefficient between $\ftides$ and $\fcomp$ is found to be $\sim\!0.46$. It is also evident from the left panel of figure \ref{fig:fcomp_vs_quantities} that protoclumps mostly occupy the upper right quadrant, that of high $\fcomp$ and positive $\ftides$, while random patches mostly occupy the bottom left quadrant, that of low $\fcomp$ and negative $\ftides$.

The right panel of figure \ref{fig:fcomp_vs_quantities} shows $\fstr$ vs. $\fcomp$ for protoclumps and random patches. Again, we can see a positive correlation between these two quantities. $\fstr\sim1$ at $\fcomp\sim0.1$ and increases to $\fstr\sim2$ at $\fcomp\sim 0.65$. It is evident that the scatter in $\fstr$ is very large, which we attribute to the noisy definition we employ for stream material. The Spearman correlation coefficient between $\fstr$ and $\fcomp$ is found to be $0.3$. Here also protoclump regions mostly occupy the high $\fcomp$ high $\fstr$ quadrant, while random patches mostly occupy the low $\fcomp$ low $\fstr$ quadrant.

{The correlations might have been stronger if we had accounted for the relevant timescales required for tides and stream-disk interactions to induce significant local convergence. However, this approach would necessitate tracking protoclumps further back in time, which is particularly challenging in our simulations due to their Eulerian framework.}

\section{Discussion}\label{sec:discussion}
\subsection{Source of gravitational tides - cosmological vs. isolated simulations}\label{sec:discussion_source_of_tides}
In figure \ref{fig:ftides_hist}, we showed that almost all of the protoclumps reside in regions where the tidal field is substantially or fully compressive, and almost all of the regions where the tidal field is fully compressive are protoclumps regions. Furthermore, in the left panel of figure \ref{fig:fcomp_vs_quantities}, we showed a positive correlation between the compressive tendency of the tidal field and the compressive tendency of the local turbulent field. A key motivation for examining the tidal field's impact on compressive turbulence in protoclumps is the distinct difference observed in \citetalias{Mandelker24} between isolated and cosmological simulations, and the previous result that the cosmological environment can cause compressive tides that drive compressive turbulence \citep{Renaud13}. It is therefore important to try and shed some light on the source of the tides in our simulations.

With the findings of several studies \citep{Renaud09,Li22} that, during major mergers, more regions of the disk experience tidal compression, mergers are a natural candidate for the source of tides. While galaxies in the VELA simulations experience some mergers during their lifetime, they usually experience around one major merger \citep{Dekel2020,Lapiner23}, not sufficient to explain our results here. Moreover, this would have led to a systematic increase in the global power in compressive modes \citep{Renaud13}, which is not observed. However, compressive tides can arise due to other sources as well. For example, from the potential of the galaxy or its host halo - in a cored spherical density profile (i.e. with a negative logarithmic slope less than unity), the tidal force in becomes fully compressive \citep{Dekel03}. In a flattened, axisymmetric system, the tidal force in the radial direction in the midplane becomes compressive if the gravitational potential satisfies $\partial^2\phi/\partial R^2 > 0$. {In general, the tidal field will be compressive along any direction in which the gravitational vector field increases with increasing distance from the region of interest}. Given the small number of major mergers in our simulations, the potential of the dark halo or the smooth component of the disk, nearby perturbations in the disk, such as clumps, spiral arms or other overdensities, minor mergers or the generally messy environment in which the galaxy resides are more likely sources for compressive tides. 

\begin{figure*}
    \centering
    \includegraphics[width=0.45\linewidth]{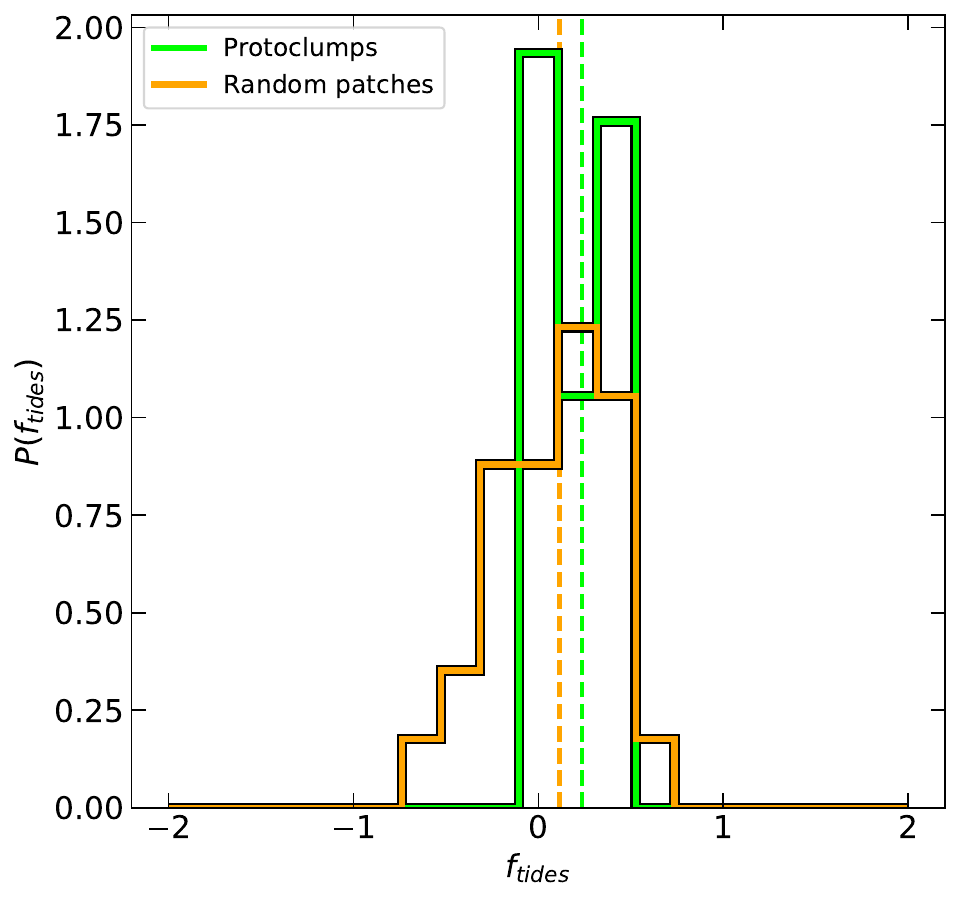}\smallskip
    \includegraphics[width=0.44\linewidth]{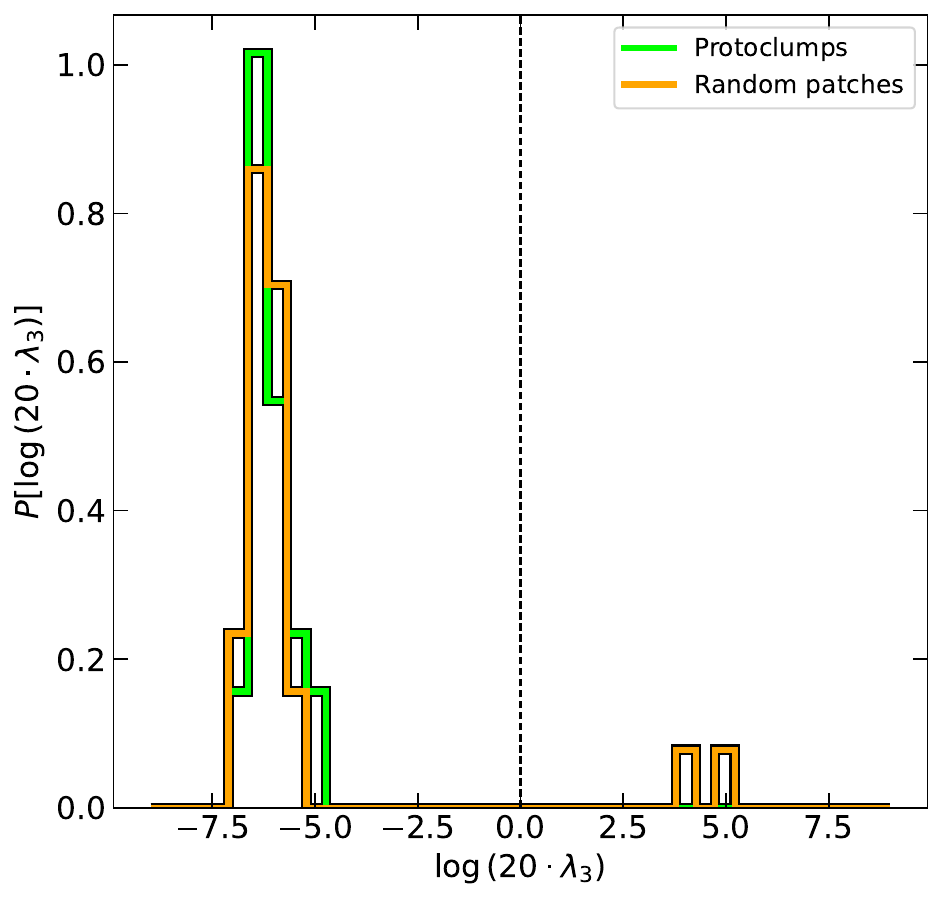}
    \caption{The same as figure \ref{fig:ftides_hist}, but for the protoclumps in the isolated galaxy simulation. We can see that the protoclumps in the isolated galaxy simulation are in regions where the tidal field is only weakly compressive, and are not much different than the random patches. Furthermore, $\lambda_3<0$ for all of the protoclump regions, as opposed to protoclump regions in the cosmological simulations.}
    \label{fig:ftides_hist_isolated}
\end{figure*}

To test the viability of in-situ induced compressive tides, we perform our tidal analysis on the same isolated simulations we analyzed in \citetalias{Mandelker24}. This simulation is an idealized simulation of a star forming galactic disk with baryonic mass of $\sim3.2\cdot10^{10}\Msun$ and initial $65\%$ gas fraction.  We refer the reader to \S2.2 of \citetalias{Mandelker24} for details on the simulation. In figure \ref{fig:ftides_hist_isolated}, we show the analog of figure \ref{fig:ftides_hist}, but for the isolated simulation. We see that the distributions of both $\ftides$ and $\lambda_3$ are qualitatively different in the isolated simulation compared to VELA. First, in the isolated simulation the distribution of $\ftides$ is very similar in protoclumps and random patches, with a median value of $\sim 0.2$. Second, all protoclumps in the isolated simulation reside in regiosn where $\lambda_3<0$, that is, no protoclump experiences fully compressive tides. 

While this is not conclusive evidence, the comparison suggests that the compressive tides in the VELA simulations likely do not originate from the potential of the halo or of the smooth disk component. However, since the clump-forming phase of the disk in the isolated simulation is relatively short, whereas in the VELA simulations it is a continuous process, we cannot rule out internal perturbations, as noted above, as potential sources of the tides. A more detailed analysis is required to identify whether the tides arise from internal or external perturbations, which is beyond the scope of this paper.

\subsection{Efficiency of stream-driven turbulence}\label{sec:discussion_efficiency}
When streams interact with a galactic disk, we expect them to drive turbulence within the disk. However, the efficiency of this stream-driven turbulence remains a topic of debate. Some studies suggest that accretion has little to no effect on disk turbulence \citep{Hopkins13}, while others argue that its impact is weak \citep{ElmegreenBurkert10}, occasionally significant \citep{Gabor13}, or even highly efficient \citep{Forbes23, Jimenez23}. If accretion-driven turbulence is indeed efficient, it could play a major role in sustaining overall disk turbulence \citep{Ginzburg22}, particularly compressive turbulence. Preliminary controlled experiments investigating streams feeding a galactic disk (Ginzburg et al., in prep) indicate that disks fed by streams tend to sustain larger values of velocity dispersions, with a strong dependence on the density of the incoming streams.

Stream density is expected to play a key role \citep{KlessenHennebelle}, as basic momentum conservation suggests that more kinetic energy is retained in the system when the densities of the colliding materials are comparable\footnote{For instance, in a simple, perfectly inelastic collision, the fraction of kinetic energy retained in the system is proportional to $\delta / (1 + \delta)$, where $\delta$ is the density contrast between the colliding materials. While this relationship may differ in fluid collisions, it highlights the significance of density contrast in such interactions.}. Since the disk is built by the incoming material, streams cannot be much less dense than the outer disk. Moreover, simulations of streams feeding massive halos show that streams grow denser as they approach the inner halo \citep{Aung24}, undergoing gravitational fragmentation along the way \citep{Mandelker18,Aung19}. Furthermore, \citet{Folini2014} found in simulations of head on collision of isothermal flows that high Mach number collisions are efficient at converting collision kinetic energy to turbulent kinetic energy, and produce a broader density distribution in the collision region.

Future work is planned to explore stream-disk interactions and the efficiency of accretion-driven turbulence, from cloud to galactic scales. These studies will also shed light on how compressive turbulence is driven in sites of stream-disk interaction.

\subsection{Definition of compressive modes of turbulence}\label{sec:discuss_definition_of_compressive_modes}
The motivation for the definition of $\fcomp$ as defined in eq. \ref{eq:fcomp_def}, is based on the 
Taylor expansion performed in eq. \ref{eq:local_decomp} (see also Appendix \ref{sec:viscous_dissipation_proof}). This definitions neglects the shearing term, which we termed $\overleftrightarrow{S}$, which is perhaps justified only for periodic or vanishing boundary conditions \citep{Kritsuk07}, both of which are clearly irrelevant for our local analysis. We can gain qualitative insights on the relevance of $\shear$ from the following argument.

The matrix $\shear$ is a symmetric matrix, and hence can be orthogonally diagonlized by three real eigenvalues, $\mu_1\geq\mu_2\geq\mu_3$. Furthermore, it is traceless, meaning that $\mu_1+\mu_2+\mu_3=0$. It therefore follows that $\mu_1\geq 0\geq \mu_3$, with $\mu_2$ somewhere in between. It means that the contribution from $\shear$ along the direction corresponding to $\mu_3$ is converging, while the contribution from $\shear$ along the direction corresponding to $\mu_1$ is diverging. Since the sum of the eigenvalues is zero, $\mu_1$ and $|\mu_3|$ can only differ at most by a factor of two. Thus, assuming $\mu_1\sim|\mu_3|$ is a reasonable approximation. We can then write the Taylor expansion of eq. \ref{eq:local_decomp} in the coordinate system determined by $\shear$ as
\begin{equation}
    \vec{v}(\vec{r}+\delta\vec{r}') \approx \vec{v} + 
    \begin{pmatrix}
        \left(1/3\lbrac{\vec\nabla\cdot\vec{v}} + \mu_1\right)\delta x' \\
        1/3\lbrac{\vec\nabla\cdot\vec{v}}\delta y' \\
        \left(1/3\lbrac{\vec\nabla\cdot\vec{v}} - \mu_1\right)\delta z'\\
    \end{pmatrix} + \frac{1}{2}\vec\omega'\times\delta\vec{r}',
\end{equation}
where primed values are the vectors in the lab frame represented in the coordinate frame of $\shear$. It is therefore evident that neglecting $\shear$ is reasonable only if $\left|\nabla\cdot\vec{v}\right| \gg \mu_1$. However, for a region to become dense enough for self gravity to become efficient, isotropic collapse is not needed. It is therefore enough that $\divr{\vec{v}<}0$ and $|\nabla\cdot\vec{v}|\gtrsim \mu_1$ so that the flow along th $x'$ will be weakly diverging, and converging flow will occur along $y'$ and $z'$ direction. 

When examining the simulations, we find that indeed usually $\mu_2\sim0$, while $\mu_1\sim|\mu_3|$. protoclump regions usually have smaller shear eigenvalues than other random patches in the disk. Furthermore, we find that $|\nabla\cdot\vec{v}| \sim 2\mu_1$ in protoclumps, while $|\nabla\cdot\vec{v}|\lesssim\mu_1$ in random patches. Alongside the sub-dominance of converging modes in random patches, we conclude that shear has a weaker effect in protoclump regions compared to random patches, which are more susceptible to shear, which prevents clump formation even when stellar feedback is weak \citep{Fensch2020}. {The increased shear can be due to the mean galactic rotation or the induction of strong shear in spiral arms. The exact cause of this shear should be explored in detailed, and is beyond the scope of this paper.}

A more robust, multi scale method is required to go beyond the Helmholtz decomposition and Taylor approximation. A natural candidate is by utilizing wavelet transforms, which are common in turbulence studies in fluid dynamics \citep{Farge92}, but seem to be less common in galactic fluid dynamics. Not only that, but a locally orthogonal Helmholtz-like decomposition algorithm using wavelets has been formalized \citep{Deriaz09}. Future work will be dedicated to performing turbulence decomposition using wavelets, with the potential for a more self-consistent quantification of compressive and solenoidal modes of turbulence.

\section{Conclusions}\label{sec:conclusions}
By analyzing cosmological simulations of violent disks at $1\!\lesssim\!z\!\lesssim4$, we studied the turbulent nature of protoclumps - regions out of which giant, star forming clumps form. Protoclump regions in the VELA cosmological simulations show local values of the Toomre-$Q$ parameter greater than unity, sometimes substantially greater \citep{Inoue16}, indicating that their subsequent gravitational collapse is not initiated by linear gravitational instabilities. An excess in compressive modes of turbulence, and in particular converging modes, can on the one hand increase the value of the velocity dispersion, and therefore increase $Q$, but on the other hand cause the local material to become dense enough for self-gravity to eventually kick in \citep{Hopkins13b}.

By extending the sample size of galaxies and protoclump regions, we strengthened the conclusion of \citetalias{Mandelker24}, finding that protoclumps are dominated by converging modes of turbulence. In our extended sample, we find that $\sim 50-70\%$ of the turbulent kinetic energy in protoclumps is in converging modes, compared to $\sim 16\%$ expected in a fully isotropic and homogeneous turbulence in equipartition. Such an excess was not found in isolated galaxies, as we have shown in \citetalias{Mandelker24}, implying that the cosmological environment or its influence on the galaxy causes its turbulence to be overly compressive.

We examine two external mechanisms for generating the excess of compressive turbulence, namely compressive tides and interactions between the disk and dense streams accreting from the cosmic web. First, the messy environment of galaxies in a cosmological setting, including merging galaxies, gives rise to gravitational tides that can be substantially compressive, and at times fully compressive \citep{Dekel03,Renaud09,Li22}. We quantify the compressiveness of the tidal field by the dimensionless quantity $\ftides$ (eq. \ref{eq:f_tides}). We find that almost all protoclump regions have $\ftides > 0$, indicating that they reside in regions where the tidal field is substantially compressive, with $\ftides \sim0.32$ on average. On the other hand, random patches in the disk have $\ftides\lesssim0$, indicating that the tidal field in these regions is substantially expansive, at least along one direction. Furthermore, we found that $\sim25\%$ of the protoclumps reside in regions where the tidal field is fully compressive, while practically no random patches are regions of fully compressive tides.

Second, high redshift massive galaxies in cosmological environments are fed by gaseous streams , typically three \citep{Danovich12,Codis18}. Upon impact, the streams can cause a strong compression due to shocks. We find that around $70\%$ of the protoclump regions are sites of stream-disk interaction, containing $\sim 2-10$ times as much stream material as the average at their galactocentric distance. Random patches, on the other hand, are more typical, with the mass in stream material in them close to the average at their galactocentric distance. We therefore conclude that protoclump regions are distinct, both in terms of experiencing substantially compressive tides and residing in sites of stream-disk interaction.

We then turn to examine how these two mechanisms correlate with the fraction of energy in compressive motion in the protoclumps and random patches in the disk. We find a positive correlation between the fraction of energy that is in compressive motion to the compressiveness of the tidal field in the region. Regions of the disk in which the tidal field is expansive typically have $\fcomp\sim0.16$, which is the value expected for a fully isotropic and homogeneous turbulence in equipartition. As the tidal field becomes more compressive, the fraction of energy in converging modes increases to $\fcomp\sim0.6-0.7$, on average. Similarly, we find a positive correlation between the intensity of the accreting streams in a given region to the fraction of energy in converging modes in the same region. Regions that are not intensely fed by streams have $\fcomp\sim 0.16$, while sites of stream-disk interaction have $\fcomp\sim0.5-0.7$.

Our results suggest that both compressive tides and stream-disk interactions can drive compressive modes of turbulence. A more detailed investigation into the sources of compressive tides is required (see Sect. \ref{sec:discussion_source_of_tides}). However, an initial comparison with an isolated simulation suggests that these tides are not caused by the potential of the dark matter halo or the smooth disk component but instead arise either from external perturbations or from internal perturbations absent during clump formation in the isolated simulations. Furthermore, the lack of tracer particles in the VELA simulations makes the stream identification crude and noisy. We plan to perform a detailed analysis of using cosmological simulations with tracer particles, in order to properly isolate accreted material from the overall gas in the disk.

The results of \citet{Inoue16}, \citetalias{Mandelker24} and the current study, suggest that there is a need for a complementary physical theory for clump formation in cosmological disk galaxies. It is evident from these works that Toomre instability does not fully describe disk fragmentation in cosmological contexts. Furthermore, the results of \citet{DuttaChowdhury24} indicate that Toomre-based models for radial mass transport \citep{DSC,KB10} generally over-predict the radial velocities observed in numerical simulations. While \citet{Hopkins13b} provide a turbulence-dependent stability threshold for $Q$ based on a statistical analysis, a more thorough analysis, based on first principles, is required. {Such a theory must take into account the balance between converging modes and solenoidal and diverging modes of turbulence. The former induces a local collapse, assisting gravity, while the latter two prevent collapse, assisting rotation and shear.}

\begin{acknowledgements}
      This work was partly supported by ISF grant 861/20, BSF-NSF grant 2023723, and by BSF-NSF grant 2023730 and by grant JWST-AR-03305.005-A. NM acknowledges support from Israel Science Foundation (ISF) grant 3061/21. DC is a Ramon-Cajal Researcher and is supported by the Ministerio de Ciencia, Innovacion y Universidades (MICIU/FEDER) under research grant PID2021122603NB-C21. The VELA simulations were performed at the National Energy Research Scientific Computing Center (NERSC) at Lawrence Berkeley National Laboratory, and at NASA Advanced Supercomputing (NAS) at NASA Ames Research Center.
\end{acknowledgements}

\bibliographystyle{aa}
\bibliography{bib}

\begin{appendix}
\section{Correspondence between the local and global decompositions}\label{app:global_vs_loca_proof}
In this appendix we show that, if the compressive and solenoidal components have proportional power spectra, the correspondence indicated in eq. \ref{eq:fcomp_global_and_local_relation} holds. We start by writing the divergence and curl using the Fourier transform
\begin{equation}
    \curl{\vec{v}} = \frac{1}{(2\pi)^{3/2}}\int i\vec{k}\times\widetilde{\vec{v}}\lbrac{\vec{k}}e^{i\vec{k}\cdot\vec{r}}d^3\vec{k},
\end{equation}
\begin{equation}
    \divr{\vec{v}} = \frac{1}{(2\pi)^{3/2}}\int i\vec{k}\cdot\widetilde{\vec{v}}\lbrac{\vec{k}}e^{i\vec{k}\cdot\vec{r}}d^3\vec{k},
\end{equation}
where $\widetilde{\vec{v}}$ is the Fourier transform of $\vec{v}$. Using Parseval's identity, we can write
\begin{equation}\label{eq:curl_parseval}
    \int\left|\curl{\vec{v}}\right|^2d^3\vec{r} = \int\left|\vec{k}\times\widetilde{\vec{v}}\right|^2d^3\vec{k},
\end{equation}
\begin{equation}\label{eq:div_parseval}
    \int\left|\divr{\vec{v}}\right|^2d^3\vec{r} = \int\left|\vec{k}\cdot\widetilde{\vec{v}}\right|^2d^3\vec{k}.
\end{equation}
Next, we can write $\widetilde{v}=\widetilde{\vec{v}}_{\rm comp} + \widetilde{\vec{v}}_{\rm sol}$ (see eq. \ref{eq:fourier_decomp}). Notice that $\widetilde{\vec{v}}_{\rm comp}\parallel \vec{k}$, while $\widetilde{\vec{v}}_{\rm sol}\perp\vec{k}$. We can therefore rewrite eqs. \ref{eq:curl_parseval} and \ref{eq:div_parseval} as
\begin{equation}
    \int\left|\curl{\vec{v}}\right|^2d^3\vec{r} = \int\left|\vec{k}\right|^2\left|\widetilde{\vec{v}}_{\rm sol}\right|^2d^3\vec{k},
\end{equation}
\begin{equation}
    \int\left|\divr{\vec{v}}\right|^2d^3\vec{r} = \int\left|\vec{k}\right|^2\left|\widetilde{\vec{v}}_{\rm comp}\right|^2d^3\vec{k}.
\end{equation}

So far, we have not made any assumption about the statistical nature of the velocity field. Next, we assume that $\left|\widetilde{\vec{v}}_{\rm sol}\right|^2 = \alpha\left|\widetilde{\vec{v}}_{\rm comp}\right|^2$, where $\alpha$ is a constant in a homogeneous turbulence. This assumption implies that the driving scale and turbulence cascade of both the compressive and solenoidal components are the same. We can therefore write
\begin{equation}
\begin{split}
    & \frac{\int\left|\divr{\vec{v}}\right|^2d^3\vec{r}}{\int\left|\divr{\vec{v}}\right|^2d^3\vec{r}+\int\left|\curl{\vec{v}}\right|^2d^3\vec{r}} = \\ 
    &\frac{\int\left|\vec{k}\right|^2\left|\widetilde{\vec{v}}_{\rm comp}\right|^2d^3\vec{k}}
    {\int\left|\vec{k}\right|^2\left|\widetilde{\vec{v}}_{\rm comp}\right|^2d^3\vec{k}+\int\left|\vec{k}\right|^2\left|\widetilde{\vec{v}}_{\rm sol}\right|^2d^3\vec{k}} = \\
    & \frac{\int\left|\vec{k}\right|^2\left|\widetilde{\vec{v}}_{\rm comp}\right|^2d^3\vec{k}}{\int\left|\vec{k}\right|^2\left|\widetilde{\vec{v}}_{\rm comp}\right|^2d^3\vec{k}+\alpha\int\left|\vec{k}\right|^2\left|\widetilde{\vec{v}}_{\rm comp}\right|^2d^3\vec{k}} = \frac{1}{1+\alpha}.
\end{split}
\end{equation}
Furthermore, again exploiting Parseval's theorem,
\begin{equation}
    \begin{split}
        & \frac{\int\limits\left|\vec{v}_{\rm comp}\right|^2d^3\vec{r}}{\int\limits\left|\vec{v}\right|^2d^3\vec{r}} = \\ 
        & \frac{\int\limits\left|\vec{v}_{\rm comp}\right|^2d^3\vec{r}}{\int\limits\left|\vec{v}_{\rm comp}\right|^2d^3\vec{r}+\int\limits\left|\vec{v}_{\rm sol}\right|^2d^3\vec{r}} = \\
        & \frac{\int\limits\left|\widetilde{\vec{v}}_{\rm comp}\right|^2d^3\vec{k}}{\int\limits\left|\widetilde{\vec{v}}_{\rm comp}\right|^2d^3\vec{k}+\int\limits\left|\widetilde{\vec{v}}_{\rm sol}\right|^2d^3\vec{k}} = \\
        & \frac{\int\limits\left|\widetilde{\vec{v}}_{\rm comp}\right|^2d^3\vec{k}}{\int\limits\left|\widetilde{\vec{v}}_{\rm comp}\right|^2d^3\vec{k}+\alpha\int\limits\left|\widetilde{\vec{v}}_{\rm comp}\right|^2d^3\vec{k}} = \frac{1}{1+\alpha}.
    \end{split}
\end{equation}

We can therefore see that the equality of eq. \ref{eq:fcomp_global_and_local_relation} holds.
\section{Viscous dissipation decomposition}\label{sec:viscous_dissipation_proof}
The viscous dissipation rate per unit volume of a compressible fluid is given by \citep{Chandrasekhar61}
\begin{equation}\label{eq:visc_diss}
    \Phi = 2\mu\lbrac{e_{ij}e_{ij} - \frac{1}{3}(e_{ii})^2}
\end{equation}
where $\mu$ is the dynamic viscosity, $e_{ij} = 1/2(\partial v_i/\partial r_j+\partial v_j/\partial r_i)$, and we have employed the summation notation. We notice that, since repeated indices are summed over,
\begin{equation}\label{eq:e_ije_ij}
e_{ij}e_{ij} = \frac{1}{4}\lbrac{\partd{v_i}{r_j}\partd{v_i}{r_j}+\partd{v_j}{r_i}\partd{v_j}{r_i}+2\partd{v_i}{r_j}\partd{v_j}{r_i}} = \frac{1}{2}\lbrac{\partd{v_i}{r_j}\partd{v_i}{r_j}+\partd{v_i}{r_j}\partd{v_j}{r_i}}.
\end{equation}
Furthermore,
\begin{equation}\label{eq:vorticity_mag}
\begin{split}
\left|\curl{\vec{v}}\right|^2 & = \epsilon_{kij}\partd{v_j}{r_i}\epsilon_{klm}\partd{v_m}{r_l} = \lbrac{\delta_{il}\delta_{jm} - \delta_{im}\delta_{jl}}\partd{v_i}{r_j}\partd{v_l}{r_m} \\
& = \partd{v_i}{r_j}\partd{v_i}{r_j} - \partd{v_i}{r_j}\partd{v_j}{r_i}.
\end{split}
\end{equation}
Here, $\epsilon_{kij}$ is the Levi-Civita symbol and $\delta_{ij}$ is the Kronecker delta. Plugging into eq. \ref{eq:e_ije_ij}, we get
\begin{equation}\label{eq:e_ije_ij_2}
    e_{ij}e_{ij} = \frac{1}{2}\lbrac{\left|\curl{\vec{v}}\right|^2 + 2\partd{v_i}{r_j}\partd{v_j}{r_i}}.
\end{equation}
Next, we write
\begin{equation}\label{eq:dvi_drj}
    \partd{v_i}{r_j}\partd{v_j}{r_i} = \partd{v_i}{r_j}\partd{v_j}{r_i} - \partd{v_i}{r_i}\partd{v_j}{r_j} + \partd{v_i}{r_i}\partd{v_j}{r_j} = \partd{v_i}{r_j}\partd{v_j}{r_i} - \partd{v_i}{r_i}\partd{v_j}{r_j} + (e_{ii})^2.
\end{equation}
Using the product rule,
\[
    \partd{v_i}{r_j}\partd{v_j}{r_i} = \partd{}{r_j}\lbrac{v_i\partd{v_j}{r_i}} - v_i\partd{^2v_j}{r_j\partial r_i},
\]
\[
    \partd{v_i}{r_i}\partd{v_j}{r_j} = \partd{}{r_j}\lbrac{v_j\partd{v_i}{r_i}}-v_j\partd{^2v_i}{r_i\partial r_j}.
\]
Subtracting, as in eq. \ref{eq:dvi_drj}, and by changing $i\leftrightarrow j$ in the second term on the right hand side of the second line, we get
\begin{equation}
    \partd{v_i}{r_j}\partd{v_j}{r_i} - \partd{v_i}{r_i}\partd{v_j}{r_j} = \partd{}{r_j}\lbrac{v_i\partd{v_j}{r_i}-v_j\partd{v_i}{r_i}}.
\end{equation}
Plugging this into \ref{eq:dvi_drj}, and then to \ref{eq:e_ije_ij_2}, we get
\begin{equation}
    e_{ij}e_{ij} = \frac{1}{2}\lbrac{\left|\curl{\vec{v}}\right|^2 + \vec{\nabla}\cdot\lbrac{\overleftrightarrow{D}\vec{v}} + 2(e_{ii})^2},
\end{equation}
where we have defined $D_{ij} = 2\partial{v_j}/\partial{r_i} - 2\lbrac{\divr{\vec{v}}}\delta_{ij}$. Wrapping everything back to eq. \ref{eq:visc_diss}, we get
\begin{equation}
    \Phi = \mu\left|\curl{\vec{v}}\right|^2 + \frac{4}{3}\mu\lbrac{\divr{\vec{v}}}^2+\mu\vec{\nabla}\cdot\lbrac{\overleftrightarrow{D}\vec{v}}
\end{equation}

Thus, the total viscous dissipation rate in a given volume $V$, is
\begin{equation}\label{eq:total_diss}
    \epsilon = \int\Phi d^3\vec{r} = \mu\int\limits_{V}\lbrac{\left|\curl{\vec{v}}\right|^2 + \frac{4}{3}\lbrac{\divr{\vec{v}}}^2}d^3\vec{r} + \mu\int\limits_{\partial V}\lbrac{\overleftrightarrow{D}\vec{v}}\cdot d^2\vec{r},
\end{equation}
where we have used the divergence theorem to turn the volume integral of $\divr{\lbrac{\overleftrightarrow{D}\vec{v}}}$ into a surface integral. From eq. \ref{eq:total_diss}, we learn that the viscous dissipation rate, and hence the turbulence dissipation rate, is affected by the curl and the divergence of the velocity field inside the volume \citep{Kida90}, but also from a surface term. This surface term vanishes only for specific boundary conditions, like periodic and vanishing at infinity. In these cases, if one thinks of the viscous dissipation rate as the energy injection rate to turbulence, one can qualitatively expect that the total energy in compressive modes of turbulence is $\propto\left|\divr{\vec{v}}\right|^2$, and the total energy in solenoidal modes is $\propto\!\left|\curl{\vec{v}}\right|^2$.
\end{appendix}

\end{document}